\documentclass[11pt]{article}

\usepackage{amsfonts}
\usepackage{appendix}
\usepackage{verbatim}

\usepackage{amsmath}
\usepackage{amssymb}
\usepackage{latexsym}
\usepackage{graphicx}
\usepackage[english]{babel}
\usepackage[font={small}]{caption}
\usepackage{mathtools}
\usepackage{geometry}                
\usepackage{epstopdf}
\usepackage{tikz-cd}
\usetikzlibrary{cd}
\usepackage{amsfonts}
\usepackage{latexsym}
\usepackage{graphicx}
\usepackage[english]{babel}
\usepackage{tikz}

\begin{document}
\input epsf

\def\p{\partial}
\def\h{{1\over 2}}
\def\be{\begin{equation}}
\def\bea{\begin{eqnarray}}
\def\ee{\end{equation}}
\def\eea{\end{eqnarray}}
\def\d{\partial}
\def\la{\lambda}
\def\eps{\epsilon}
\def\b{\bigskip}
\def\m{\medskip}

\newcommand{\newsection}[1]{\section{#1} \setcounter{equation}{0}}

\def\q{\quad}

\def\h{{1\over 2}}
\def\t{\tilde}
\def\r{\rightarrow}
\def\nn{\nonumber\\}

\let\p=\partial

\newcommand\blfootnote[1]{%
  \begingroup
  \renewcommand\thefootnote{}\footnote{#1}%
  \addtocounter{footnote}{-1}%
  \endgroup
}

\begin{flushright}
\end{flushright}
\vspace{20mm}
\begin{center}
{\LARGE Lifting of  level-1 states  in the D1D5 CFT}
\\
\vspace{18mm}
{\bf   Bin Guo$^1$\blfootnote{$^{1}$guo.1281@osu.edu} and Samir D. Mathur$^2$\blfootnote{$^{2}$mathur.16@osu.edu}
\\}
\vspace{10mm}
Department of Physics,\\ The Ohio State University,\\ Columbus,
OH 43210, USA\\ \vspace{8mm}

\vspace{8mm}
\end{center}

\vspace{4mm}

\thispagestyle{empty}
\begin{abstract}

The D1D5 CFT has a large set of states that are supersymmetric at the `free' orbifold point in moduli space. When we perturb away from this point, some of these states join into long multiplets and lift in energy, while others remain supersymmetric. The count of unlifted states can be bounded below by an index, but the index does not yield the pattern of lifting; i.e., which states join into a long multiplet and how much this multiplet lifts. In this paper we consider the simple case of the D1D5 CFT where the orbifold CFT is a sigma model with targets space $(T^4)^2/S_2$ and consider states at energy level $1$. There are $2688$ states at this level. The lifted states form  a triplet of long multiplets, and we compute their lift at second order in perturbation theory. Half the members of the long multiplet are in the untwisted sector and half are in the twisted sector. This and other similar studies should help in the understanding of   fuzzball states that describe extremal holes, since CFT sectors with low twist describe shallow throats in the dual gravity solution while sectors with high twist describe deep throats.

\vspace{3mm}

\end{abstract}
\newpage

\setcounter{page}{1}

\numberwithin{equation}{section} 

\tableofcontents

\section{Introduction}

The D1D5P system provides a very useful instance of an extremal black hole in string theory \cite{sv, cm, dmcompare, maldastrom}. The D1 and D5 branes form a bound state whose dynamics can be given as an effective $1+1$ dimensional conformal field theory. The momentum charge P is given by the difference in energy between the left moving and right moving excitations in this CFT. Our interest is in the extremal states of this CFT; i.e., those where the right moving sector is in its ground state. Such states correspond to microstates of the extremal black hole, and preserve 1/8 of the supersymmetries of the string theory.  

The CFT has a `free point' which is given by a $1+1$ dimensional sigma model whose target space is an orbifold \cite{Vafa:1995bm,Dijkgraaf:1998gf,orbifold2,Larsen:1999uk,Arutyunov:1997gt,Arutyunov:1997gi,Jevicki:1998bm,David:2002wn}. The orbifold theory consists $N$ copes of a $c=6$ CFT, joined up in different `twist sectors'.  In each twist sector the excitations are just given by free left and right moving bosons and fermions, with an overall symmetry condition to enforce the orbifold symmetry. At this orbifold point,  any state with no right moving oscillator excitations is extremal. 

The situation changes as we deform the theory away from the orbifold point. Sets of extremal states can join up into larger multiplets and lift to higher energies, leaving a smaller set of states that remain extremal. The latter set is the set of microstates of the extremal black hole. Thus we are interested in finding the pattern of lifting. In other words, we are interested in the following questions. Which of the extremal states at the orbifold point remain extremal? For the states that lift, which ones join into long multiplets?  For these lifted states,  how much is the lift in energy?

Let us first recall some results that are known in this direction. The {\it count} of states that remain unlifted is given by an index. This index was computed in \cite{sv} for the case where the compactification is $K3\times S^1$ and in \cite{mms} for the compactification $T^4\times S^1$.  Looking at the expression for the index in \cite{mms}, one notes that if the numbers of D1, D5, P charges are taken to be co-prime, then the index can be reproduced by assuming that all states in the maximally twisted sector of the CFT are unlifted, and all states in the other twist sectors are lifted. The actual unlifted states do not, of course, have to be in the maximally twisted sector: the index gives a (weighted) count of unlifted states, without telling us what these states look like. We will find that the lifted states form multiplets whose members lie in sectors with different twists, including the sector with maximal twist.  Understanding the twist sectors is important for a physical picture of the extremal hole. States in highly twisted sectors correspond to gravity states with deep throats, while states in sectors with low twist describe shallow throats. 
(For constructions of fuzzballs see for example \cite{fuzzballs_i,fuzzballs_ii,fuzzballs_iii,fuzzballs_iv,fuzzballs_v}.)

In \cite{gn} the lift was computed, in an approximation scheme,  for the situation where  most of the CFT copies are in the untwisted sector and one set is in a twisted sector. Low energy excitations of this sector can be mapped, in the gravity dual,  to strings in an $AdS_3\times S^3\times T^4$ spacetime. Apart from a small set of states in the graviton multiplet, these string states are all lifted. On the other hand we know from the index computation of \cite{mms} that if we go to sufficiently high energies and twists to describe black hole states, then a large number of states must remain {\it unlifted}: the index of \cite{mms} agrees with the Bekenstein entropy of extremal holes for large charges. It would be very interesting to understand better what properties of the highly excited states makes them remain `unlifted'. 

Our steps and results are as follows:

\b

(i) We consider a CFT with $N=2$; this means that the product of the number $N_1$ of D1 branes and $N_5$ of D5 branes is $n_1n_5\equiv N=2$. Further, we consider the lowest nontrivial amount of momentum charge: $P=1$. Even with these choices, the number of extremal states at this level in the orbifold CFT is  $2688$, which is a largish number. Our goal is to find the pattern of lifting for all the states at this level in this $N=2$ CFT.

\b

(ii) First we organize the states in multiplets of symmetry charges at the orbifold point. We note that states related by these symmetry charges will have the same lift. We find the lowest weight states under these symmetries, and henceforth concentrate on the lifting of these lowest weight states.

\b

(iii) The deformation operator gives a supercharge that groups states into multiplets as they lift off the orbifold point. We find  there is one triplet of  lifted multiplets. For these multiplets, we find the lift is
\be
\delta E= \lambda^2  \pi^2
\ee
where $\lambda$ is the coupling giving the deformation off the orbifold point.

\b

Before proceeding,  we note that there are many earlier works that study conformal perturbation theory, the lifting  of the states, the acquiring of anomalous dimensions, and the issue of operator mixing,   in particular in the context of the D1-D5 CFT;  see for example \cite{Avery:2010er,Avery:2010hs,Pakman:2009mi,Burrington:2012yq,Burrington:2014yia,Burrington:2017jhh,Carson:2016uwf}. Also, for more computations in  conformal perturbation theory in two and higher dimensional CFTs see, e.g.  \cite{kadanoff,Dijkgraaf:1987jt,Cardy:1987vr,Kutasov:1988xb,Eberle:2001jq,Gaberdiel:2008fn,Berenstein:2014cia,Berenstein:2016avf,gz,hmz,Keller:2019suk,Keller:2019yrr}.

\section{The D1D5 CFT}

In this section, we summarize some properties of the D1D5 CFT at the orbifold point and the deformation operator that we will use to perturb away from the orbifold point. For more details, see \cite{Avery:2010er,Avery:2010hs}.

Consider type IIB string theory, compactified as
\be
M_{9,1}\rightarrow M_{4,1}\times S^1\times T^4.
\label{compact}
\ee
Wrap $N_1$ D1 branes on $S^1$, and $N_5$ D5 branes on $S^1\times
T^4$. The bound state of these branes is described by a field
theory. We think of the $S^1$ as being large compared to the $T^4$, so
that at low energies we look for excitations only in the direction
$S^1$.  This low energy limit gives a conformal field theory (CFT) on
the circle $S^1$.

 It has been conjectured that we can move in the moduli space of couplings in the string theory to a point called the `orbifold point' where the CFT is particularly simple. At this orbifold point the CFT is
a 1+1 dimensional sigma model. We will work in the Euclidized theory, where
the base space is a cylinder spanned by the coordinates 
\be
\tau, \sigma: ~~~0\le \sigma<2\pi, ~~~-\infty<\tau<\infty
\ee
The target space of the sigma model is the `symmetrized product' of
$N_1N_5$ copies of $T^4$,
\be
(T^4)^{N_1N_5}/S_{N_1N_5},
\ee
with each copy of $T^4$ giving 4 bosonic excitations $X^1, X^2, X^3,
X^4$. It also gives 4 fermionic excitations, which we call $\psi^1,
\psi^2, \psi^3, \psi^4$ for the left movers, and $\bar\psi^1,
\bar\psi^2,\bar\psi^3,\bar\psi^4$ for the right movers. The fermions can be
antiperiodic or periodic around the $\sigma$ circle. If they are
antiperiodic on the $S^1$ we are in the Neveu-Schwarz (NS) sector, and
if they are periodic on the $S^1$ we are in the Ramond (R)
sector. The central charge of the theory with fields
$X^i, \psi^i, ~i=1\dots 4$ is $c=6$. 
The total central charge of the entire system is thus 
\be
c=6 N_1N_5\equiv 6N
\ee

\subsection{Symmetries of the CFT}

The D1D5 CFT has $(4,4)$ supersymmetry, which means that we have
$\mathcal{N}=4$ supersymmetry in both the left and right moving
sectors.
This leads to a superconformal ${\cal N}=4$ symmetry in both
the left and right sectors, generated by operators $L_{n}, G^\pm_{r},
J^a_n$ for the left movers and $\bar L_{n}, \bar G^\pm_{r}, \bar
J^a_n$ for the right movers. The full symmetry is actually larger; it is the contracted large $\mathcal{N}=4$ superconformal symmetry \cite{mms,Sevrin:1988ew}. The algebra generators and commutators are given in Appendix~\ref{commutators}. 

Each ${\cal N} = 4$ algebra has an internal R symmetry group
$SU(2)$, so there is
a global symmetry group $SU(2)_L\times SU(2)_R$.  We denote the
quantum numbers in these two $SU(2)$ groups as
\be
SU(2)_L: ~(j, m);~~~~~~~SU(2)_R: ~ (\bar j, \bar m).
\ee
In the geometrical setting of the CFT, this symmetry arises from the
rotational symmetry in the 4 space directions of $M_{4,1}$:  we have $SO(4)_E\simeq SU(2)_L\times SU(2)_R$.
Here the subscript $E$ stands for `external', which denotes that these
rotations are in the noncompact directions.  We have another $SO(4)$ symmetry in the four directions
of the $T^4$. This symmetry we call $SO(4)_I$ (where $I$ stands for
`internal'). This symmetry is broken by the compactification of the
torus, but at the orbifold point it still provides a useful organizing
principle. We write $SO(4)_I\simeq SU(2)_1\times SU(2)_2$.
We use spinor indices $\alpha, \bar\alpha$ for $SU(2)_L$ and $SU(2)_R$
respectively. We use spinor indices $A, \dot A$ for $SU(2)_1$ and
$SU(2)_2$ respectively.

The 4 real fermions of the left sector can be grouped into complex
fermions $\psi^{\alpha A}$. The right fermions have indices $\bar{\psi}^{\bar\alpha  A}$. The bosons $X^i$ are a vector in the
$T^4$. One can decompose this vector into the $(\h, \h)$  representation of $SU(2)_1\times SU(2)_2$, which gives  scalars $X_{A\dot A}$.

\subsection{Deformation of the CFT}

The deformation of the CFT off the orbifold point is given by adding a deformation operator $D$ to the action
\be
S\r S+\lambda \int d^2 z D(z, \bar z)
\ee
where $D$ has conformal dimensions $(h, \bar h)=(1,1)$. A choice of $D$ which is a singlet under all the symmetries at the orbifold point is
\be\label{D 1/4}
D=\frac{1}{4}\epsilon^{\dot A\dot B}\epsilon_{\alpha\beta}\epsilon_{\bar\alpha \bar\beta} G^{\alpha(0)}_{\dot A, -\h} \bar G^{\bar \alpha(0)}_{\dot B, -\h} \sigma^{\beta \bar\beta}
\ee
where $\sigma^{\beta\bar\beta}$ is a twist operator of rank $2$ in the orbifold theory. Here $ G^{(0)}$ and $\bar G^{(0)}$ are the left and right moving supercharge operators at the orbifold point.

\section{Computation of the lift using the Gava-Narain method}

We are interested in finding states which have well defined scaling dimensions, and the values of these dimensions, as we move away from the orbifold point. We will work in the Ramond sector.  The ground states in this sector has dimensions $(h, \bar h)=({c\over 24}, {c\over 24})$.  At level $n$, we have a set of states with dimensions 
\be
(h, \bar h)=({c\over 24}+n, {c\over 24})
\label{sthree}
\ee
Let these states be labelled by  indices $a, b, \dots$, and written as $\Big|O^{(0)}_{a}\Big\rangle$ etc. (We will be interested in level $n=1$ in this paper.)

 It turns out that while such states receive corrections at first order in $\lambda$, the dimensions get corrections only starting at $O(\lambda^2)$.   The computation involves pulling down two copies of the deformation operator $D$ from the action, and then integrating the positions of these two $D$ operators.  We first compute the matrix elements
\bea\label{Xamp}
X_{ba}(T)=\Big\langle O^{(0)}_{b}\left(\frac{T}{2}\right)\Big|\left(\int d^{2}w_{1}D(w_{1},\bar w_{1})\right)\left(\int d^{2}w_{2}D(w_{2},\bar w_{2})\right)\Big|O^{(0)}_{a}\left(-\frac{T}{2}\right)\Big\rangle
\eea
Then we compute the matrix
\bea\label{lift matrix}
E^{(2)}_{ba}=\lim_{T\rightarrow \infty}-\frac{\lambda^2}{2T}e^{E^{(0)}T}X_{ba}(T)
\eea
where $E^{(0)}$ is the energy of the states $|O^{(0)}_{a}\rangle$ at  the orbifold point.  The eigenstates of this matrix then give the linear combinations of the $|O^{(0)}_{a}\rangle$ which have definite dimensions and the eigenvalues give the lift in energy of the corresponding states. 

Such $O(\lambda^2)$ corrections were computed for some simple states in \cite{gz,hmz}. In general the computation of a correlation functions with deformation operators involves going to a covering space where the effect of the twists is undone, and one gets a correlator of operators not involving twists on this covering space. But the covering space can be a sphere in some cases, and a torus in other cases. While correlators on a sphere are easy to compute, they can be difficult to find on a torus. (A central reason for this difficulty is that the correlators on the covering space can involve spin fields. On a sphere we can remove these spin fields by spectral flows, but it is not clear how to do this on a higher genus surface.)

If we cannot explicitly compute the amplitudes (\ref{Xamp}), then how can we find the lifting? 
In \cite{gn} Gava and Narain gave a method by which amplitudes like  (\ref{Xamp}) could be written as modulus squared of amplitudes involving just {\it one} twist. Computing these one-twist amplitudes always gives a covering space that is a sphere, so the computation is straightforward. 
 
In \cite{Guo:2019pzk} this proposal of \cite{gn} was studied in detail. Let us recall the results of this study. We find
\be
\epsilon_{\dot A\dot B}\epsilon^{\bar \alpha\bar \beta}E^{(2)}_{ba}=2\lambda^2\Big\langle O^{(0)}_{b}\Big|\Big\{  \bar G^{\bar\alpha(P)}_{\dot A,0},  \bar G^{\bar \beta(P)}_{\dot B,0} \Big\}\Big|O^{(0)}_{a}\Big\rangle
\ee
The operators $\bar G^{\bar\alpha(P)}_{\dot A,0}$ will be explained below. We will refer to the matrix  $E_{ba}^{(2)}$ as the lifting matrix $E^{(2)}$.

From the above relation we see that  the lifting matrix (\ref{lift matrix}) can be written using either of  the following two equivalent  expressions
\bea\label{liftmatrix 1}
E^{(2)}_{ba}=2 \lambda^2   
\Big\langle O^{(0)}_{b}\Big|    \Big\{  \bar G^{+(P)\dagger}_{+,0},  \bar G^{+(P)}_{+,0} \Big\} \Big|O^{(0)}_{a}\Big\rangle
=2 \lambda^2   
\Big\langle O^{(0)}_{b}\Big|    \Big\{  \bar G^{+(P)\dagger}_{-,0},  \bar G^{+(P)}_{-,0} \Big\} \Big|O^{(0)}_{a}\Big\rangle
\eea
Note that all operators in these expressions are at the {\it same} time $\tau$. This is in contrast to (\ref{Xamp}) where the operators are clearly at different times $\tau_1, \tau_2$. The operators $\bar G^{\bar\alpha(P)}_{\dot A,0}$ are given by
\be\label{GN p}
\bar G^{\bar \alpha (P)}_{\dot A,0}= \mathcal P\int_{0}^{2\pi} \frac{d\sigma}{2\pi} (\pi G^{+}_{\dot A,-\frac{1}{2}}\sigma^{-\bar \alpha}(\tau,\sigma))  \mathcal P
\ee

Here the  operator $\mathcal P$ is a projection operator, which projects any state to the subspace spanned by the unperturbed states $|O^{(0)}_{a}\rangle$ which have the dimensions (\ref{sthree}). 
The operator 
\be
\mathcal P (\pi G^{+}_{\dot A,-\frac{1}{2}}\sigma^{-\bar \alpha}(\tau,\sigma))  \mathcal P
\ee
does not depend on $\tau$ or $\sigma$, since the projection operators $\mathcal P$ ensures that it acts between states of the same dimension. Thus (\ref{GN p}) can be simplified to
\bea\label{GN p s}
\bar G^{\bar \alpha (P)}_{\dot A,0}= \pi \mathcal P G^{+}_{\dot A,-\frac{1}{2}}\sigma^{-\bar \alpha} \mathcal P
\eea
Note that when we apply this operator to a state with dimension (\ref{sthree}), we can drop the projection operator $\mathcal P$ on the right since it will act as   the identity.

Further, it was noted in  \cite{Guo:2019pzk}, that the operators $\bar G^{\bar \alpha (P)}_{\dot A,0}$  give the supersymmetric structure of long multiplets. At the orbifold point the states can be grouped into short multiplets. As we deform away from the orbifold point, four of these short multiplets can join into a long multiplet and lift. The structure of this long multiplet is indicated in the following diagram:
\be\label{multiplets diagram}
\begin{tikzcd}
                                                & \phi_{+} \arrow[dr,"\bar G^{+(P)}_{-,0}"]\\
\phi \arrow[ur, "\bar G^{+(P)}_{+,0}"]  \arrow{dr}[swap]{\bar G^{+(P)}_{-,0}}& &  \phi_{+-}
\\
                                                & \phi_{-} \arrow{ur}[swap]{\bar G^{+(P)}_{+,0}}
\end{tikzcd}
~~~~~~~~~
\begin{tikzcd}
                                                & \arrow{dl} [swap]{\bar G^{-(P)}_{-,0}} \phi_{+} \\
\phi   & &  \arrow{ul}[swap]{\bar G^{-(P)}_{+,0}} \phi_{+-} \arrow{dl}{\bar G^{-(P)}_{-,0}}
\\
                                                & 
                    \arrow{ul}{\bar G^{-(P)}_{+,0}} \phi_{-} 
\end{tikzcd}
\ee
The state $\phi$ is at the bottom of this long multiplet. Note that $\phi$  is a member of a short multiplet created by operators that are not depicted in the figure.  The operators $\bar G^{+(P)}_{+,0}$ and $\bar G^{+(P)}_{-,0}$ play the role of the two raising operators which take us to states $\phi_+, \phi_-$ which are members of two other short multiplets.  Acting with both these raising operators takes us to the short multiplet represented by the state $\phi_{+-}$.  We can move along this multiplet in the reverse direction using the lowering operators $\bar G^{-(P)}_{+,0}$ and $\bar G^{-(P)}_{-,0}$. 

Suppose we have diagonalized the matrix $E^{(2)}$ given in  eq. (\ref{liftmatrix 1}). Let  $|O^{(0)}\rangle$ be an eigenstate of this matrix. Let the corresponding eigenvalue,  which gives the lift of this operator, be called $E^{(2)}_{O}$. Then from (\ref{liftmatrix 1}) we find that $E^{(2)}_O$ can be written as a sum of modulus-squared terms
\be\label{lifting norm}
E^{(2)}_O=2 \lambda^2   
\Big(\Big|\bar G^{+(P)}_{+,0} |O^{(0)}\rangle\Big|^2+\Big|\bar G^{-(P)}_{-,0} |O^{(0)}\rangle\Big|^2\Big)
=2 \lambda^2   
\Big(\Big|\bar G^{+(P)}_{-,0} |O^{(0)}\rangle\Big|^2+\Big|\bar G^{-(P)}_{+,0} |O^{(0)}\rangle\Big|^2\Big)
\ee

In the long multiplet described in (\ref{multiplets diagram}) each of the four states $\phi$, $\phi_{+}$, $\phi_{-}$ and $\phi_{+-}$ have the following property:  if it can be raised by $\bar G^{+(P)}_{+,0}$, then it will be annihilated by the  $\bar G^{-(P)}_{-,0}$; conversely, if it can be lowered by $\bar G^{-(P)}_{-,0}$ then it will be annihilated by  the $\bar G^{+(P)}_{+,0}$.  The same holds for the raising operators $\bar G^{+(P)}_{-,0}$ and the lowering operators $\bar G^{-(P)}_{+,0}$. Thus in each of the two expressions in (\ref{lifting norm}), only one of the two terms is nonzero. 

In summary, one can get the lifting and corresponding eigenstates by diagonalizing the lifting matrix $E^{(2)}$ (\ref{liftmatrix 1}). 
Four short multiplets join into a long multiplet as shown in eq. (\ref{multiplets diagram}) and the lifting can be calculated from (\ref{lifting norm}).

\section{The global zero mode multiplets}\label{s global multiplet}

The CFT has left and right moving symmetries generated by its  chiral algebra. These symmetries remain true for all values of the coupling. Thus states related by these symmetries will have the same lift $E^{(2)}$. We would like to group states that are related by these symmetries, so that we may reduce the number of independent lifting computations that we have to perform. We will call the operators giving these symmetries `global zero modes' since they act on all the $N$ copies of the $c=6$ CFT.

 In  subsection \ref{algebra and character}, we introduce the subalgebra of global modes that we will use. We will build characters from the representations of these subalgebras; this will help us count all the states  within a multiplet.  In  subsection \ref{pf and counting}, we count the number of multiplets by taking the partition function of the orbifold theory and writing it in terms of the characters.  In subsection \ref{N=(1,1) multiplets} and \ref{N=(2) multiplets}, we construct the lowest weight states of these multiplets explicitly for the case of two singly wound copies of the $c=6$ CFT and for the case of  one doubly wound copy.

\subsection{The subalgebra and the characters of the multiplets}\label{algebra and character}

We consider the subalgebra formed by  the following left and right moving global zero modes
\be\label{global modes for chara}
d_{0}^{\alpha A}~~~~~J^{a}_{0} ~~~~~G^{\alpha}_{\dot A,0}~~~~~\bar d_{0}^{\bar \alpha A}
\ee
where the expression for these modes are given in (\ref{global modes app}).
We will now note that  these global modes
commute with the operators  $\bar G^{\bar \alpha(P)}_{\dot A,0}$ that act as raising and lowering operators in the long multiplet. 

First note that all these zero modes (\ref{global modes for chara}) commute with the projection operator $\mathcal P$, since zero modes do not change the dimension of an operator. Thus we must consider the commutation of the modes (\ref{global modes for chara}) with $G^{+}_{\dot A,-\frac{1}{2}}\sigma^{-\bar \alpha}$.  

First consider the left moving zero modes. Then we must consider their commutation with $G^{+}_{\dot A,-\frac{1}{2}}\sigma^{-}$.   The relation (\ref{d D zero}) gives
\bea\label{d D zerop}
\{d^{\alpha A}_{n}, G^{+}_{\dot B, -\frac{1}{2}}\sigma^{-}\}
=0
\eea
Thus we find  
\be
\{d_{0}^{\alpha A}, \bar G^{\bar \alpha(P)}_{\dot A,0}\}=0
\ee

Since $G^{+}_{\dot A,-\frac{1}{2}}\sigma^{-}$ carries no charge under $J^{a}_{0}$, we have
\be
[J^{a}_{0}, G^{+}_{\dot B, -\frac{1}{2}}\sigma^{-}]=0
\ee
Thus we get
\be
[J^{a}_{0}, \bar G^{\bar \alpha(P)}_{\dot A,0}]=0
\ee

Now we note that
\be
[G^{\alpha}_{\dot A,0}, G^{+}_{\dot B, -\frac{1}{2}}\sigma^{-}]=\epsilon_{\dot A\dot B}\p\sigma^{\alpha}
\ee
Note that in (\ref{GN p}) there is an integral over $\sigma$ which will make any total derivative vanish. Thus we get
\be
[G^{\alpha}_{\dot A,0}, \bar G^{\bar \alpha(P)}_{\dot A,0}]=0
\ee

Now consider the right moving part of $G^{+}_{\dot A,-\frac{1}{2}}\sigma^{-\bar \alpha}$, which is given by the operator  $\bar\sigma^{\bar \alpha}$. The relation  (\ref{right d zeromode sigma}) says
\be\label{right d zeromode sigmap}
[\bar d^{\bar \alpha A}_{0}, \bar \sigma^{\bar \beta}(0)]
=0
\ee
Thus we get
\be
\{\bar d_{0}^{\bar \alpha A},\bar G^{\bar \alpha(P)}_{\dot A,0}\}=0
\ee
Thus we find that the global modes (\ref{global modes for chara}) commute with the operators $\bar G^{\bar \alpha(P)}_{\dot A,0}$.

We define a short multiplet as the multiplet generated by these modes in the space of states with dimension (\ref{sthree}). The reason to use these modes is following. 
We show that these global modes
commute with the raising and lowering operators $\bar G^{\bar \alpha(P)}_{\dot A,0}$, which join four states into a long multiplet and lift. Thus all the states in a short multiplet have the same lift. 
Looking at the lowest weight state in a short multiplet will therefore suffice to give us the lifting.

Since the operators $\bar G^{\bar \alpha(P)}_{\dot A,0}$  carry $\bar J^3_0$ charge, they do not commute with the $\bar J^a_0$. Thus we do not include the $\bar J^a_0$ operators in the subalgebra that we are using.

The relevant commutation relations for the left movers are
\bea
\{G^{+}_{+,0},-G^{-}_{-,0}\}=\{G^{+}_{-,0},G^{-}_{+,0}\}=L_{0}-\frac{c}{24}\nn
\{d^{++}_{0},-d^{--}_{0}\}=\{d^{+-}_{0},d^{-+}_{0}\}=1\nn
{[}J_{0}^{3},G^{\pm}_{\dot A,0}{]}=\pm\frac{1}{2}G^{\pm}_{\dot A,0}\nn
{[}J_{0}^{3},d^{\pm A}_{0}{]}=\pm\frac{1}{2}d^{\pm A}_{0}
\eea

and for the right movers are
\bea
\{\bar d^{++}_{0},-\bar d^{--}_{0}\}&=&\{\bar d^{+-}_{0},\bar d^{-+}_{0}\}=1\nn
{ [} \bar J_{0}^{3},\bar d^{\pm A}_{0}{]}&=&\pm\frac{1}{2}\bar d^{\pm A}_{0}
\eea
The lowest weight states of the irreducible representation of this subalgebra are defined by
\be\label{lowest states L}
d^{-A}_{0}|\phi\rangle=G^{-}_{\dot A,0}|\phi\rangle=J^{-}_{0}|\phi\rangle=\bar d^{-A}_{0}|\phi\rangle=0
\ee
with charges
\be\label{lowest states R}
J^{3}_{0}|\phi\rangle=-j|\phi\rangle~~~~~~
\bar J^{3}_{0}|\phi\rangle=\bar j_{3}|\phi\rangle
\ee 

The characters  are defined as follows:
\be\label{global character}
\chi_{j\bar j_3}(y,\bar y)={\rm Tr}_{j\bar j_3}(-1)^{2J^{3}_{0}-2\bar J^{3}_{0}} y^{2J^{3}_{0}} \bar y^{2 \bar J^{3}_{0}}
\ee
where the trace is over states in the irreducible representation. We compute these characters in Appendix \ref{app chara}. The operators of the subalgebra act independently on the left and right, so the characters have a factorized form
\be
\chi_{j\bar j_3}(y,\bar y)=\chi^L_{j}(y)\chi^{R}_{\bar j_3}(\bar y)
\ee
We find
\be
\chi^L_{j}(y)
=\chi_{j-1}(-y)(y^{1/2}-y^{-1/2})^4
\ee
and
\be\label{global character right}
\chi^{R}_{\bar j_{3}}(\bar y)
=-(-\bar y)^{2\bar j_{3}+1}(\bar y^{1/2}-\bar y^{-1/2})^2
\ee
where
\be
    \chi_{j}(y)=y^{2j}+y^{2j-2}+\ldots+ y^{-2j}=\frac{y^{2j+1}-y^{-2j-1}}{y-y^{-1}}
\ee

\subsection{The partition function and the counting of multiplets}\label{pf and counting}

In this subsection, we will recall the partition function of the orbifold  CFT. We will then express this partition function in terms of the characters found above. This will allow us to find the number of multiplets of the subalgebra at our chosen level $n_p=1$.  

The partition function for a single $c=6$ copy of the CFT  is defined as
\be\label{pf single copy}
Z={\rm Tr} (-1)^{2J^{3}_{0}-2\bar J^{3}_{0}}q^{L_{0}}\bar q^{\bar L_{0}} y^{2J^{3}_{0}} \bar y^{2 \bar J^{3}_{0}}\equiv\sum_{h,\bar h,j^3,\bar j^3} c(h,\bar h,j^{3},\bar j^{3})q^{h}\bar q^{\bar h} y^{2j^3} \bar y^{2\bar j^3}
\ee
For the case where the target space is $T^4$, we consider states without $U(1)$ charges coming from the translational symmetries along the four directions of the $T^4$. For this case, one finds
\be
Z(T^4)=\left(\frac{\theta_{1}}{\eta}\right)^{2}\frac{1}{\eta^4} \overline{\left(\frac{\theta_{1}}{\eta}\right)^{2}\frac{1}{\eta^4}}
\ee
where
\bea
\theta_{1}&=&i(y^{1/2}-y^{-1/2})q^{1/8}\prod_{n=1}^{\infty}(1-q^n)(1-y q^n)(1-y^{-1}q^n)\nn
\eta&=&q^{1/24}\prod_{n=1}^{\infty}(1-q^{n})
\eea
Using the above we can find the partition function for the case where the target space of the $1+1$ dimensional CFT is the symmetric product
 $Sym^{k}(T^4)$.  The partition function for a symmetric product target space
 $Z(Sym^{k}(X))$ is given by 
\be\label{partition function}
\mathcal Z (p,q,\bar q, y, \bar y)=\sum^{\infty}_{k=0}p^{k}Z(Sym^{k}(X))= \prod_{n=1}^{\infty} \prod'_{h,\bar h, j^3,\bar j^3} \frac{1}{(1-p^{n}q^{h/n}\bar q^{\bar h/n} y^{2j^3} \bar y^{2\bar j^3})^{c(h,\bar h,j^{3},\bar j^{3})}}
\ee
where $\prod'_{h,\bar h, j^3,\bar j^3}$ is restricted so that $(h-\bar h)/n$ is an integer. The $c(h,\bar h,j^{3},\bar j^{3})$ are the degeneracies appearing in the partition function for the CFT with a single copy of the space $X$. 
We write 
\be\label{expand p f}
\mathcal Z (p,q,\bar q, y, \bar y)=\sum_{N,h,\bar h,j^{3},\bar j^{3}} c(N,h,\bar h,j^{3},\bar j^{3})p^{N} q^{h}\bar q^{\bar h} y^{2j^3} \bar y^{2\bar j^3}
\ee
This yields the degeneracies $c(N,h,\bar h,j^{3},\bar j^{3})$ of the states with total winding of the effective string $N$ and  quantum number $h,\bar h, j^{3},\bar j^{3}$. 

We will be working with the case $N=2$. There are two different twist sectors for this $N$: the case of two singly wound copies which we call $N=(1,1)$, and the case of a single doubly wound copy which we call $N=(2)$. The expression (\ref{expand p f}) gives the count of states where the contribution of both these twist sectors have been added. We are however interested in obtaining the count of states separately in these two different twist sectors. It turns out that with a little effort we can separate the two contributions in the expression (\ref{expand p f}). We will do that in what follows. 

\subsection{Global zero mode multiplets for two singly wound copies at level one}\label{N=(1,1) multiplets}

In this subsection, we will apply the counting procedure discussed above to get the number of global zero mode multiplets at level $1$ for the case $N=(1,1)$. We will also  construct the lowest weight states of these multiplets explicitly.

As noted above, from (\ref{expand p f}), we can get the count of multiplets for a given value of $N$. We are however interested in the contribution to this count where the winding is given by $N=(1,1)$. We can extract this contribution as follows. In (\ref{partition function}), let us  restrict the  product to terms  with $n=1$  and collect all the terms with dependence $p^2$. In this way, we get the contribution  from states in the $N=(1,1)$ winding sector, and not from states in the $N=(2)$ winding sector. We find
\be
Z(N=(1,1);h=1,\bar h=0)=-2(y^{1/2}-y^{-1/2})^6(\bar y^{1/2}-\bar y^{-1/2})^4
\ee
By looking at the characters derived above, we find by inspection that this function can be expressed as
\be\label{11 sector chara}
Z(N=(1,1);h=1,\bar h=0)=\left(2\chi^{L}_{j=3/2}(y)+4\chi^{L}_{j=1}(y)\right)\left(\bar \chi^{R}_{\bar j_{3}=-1}(\bar y)+2\bar \chi^{R}_{\bar j_{3}=-1/2}(\bar y)+\bar \chi^{R}_{\bar j_{3}=0}(\bar y)\right)
\ee
Note that such a $Z$ does not in general have to have a form that is factorized between the left and right sectors; this is because the symmetry requirements of the orbifold theory apply to the full state and not separately to the left and right sectors.  in the present case $Z$ just happens to have a factorized form. 

Let us now look at the actual lowest weight states of the characters. By applying the conditions for the three left sector operators in (\ref{lowest states L}) we find that $d_{-1}^{-A(1)}|0^{-}_{R}\rangle|0^{-}_{R}\rangle$ is a lowest weight state for the left sector. The representation has $j=3/2$. The two choices $A=+, -$ give us two such representations. It may seem that we can get another two representations using the second copy of the CFT; i.e., from the states  $d_{-1}^{-A(2)}|0^{-}_{R}\rangle|0^{-}_{R}\rangle$, for a total of $4$ left representations with $j=3/2$. But such is not the case. The orbifold symmetry condition forces us to consider the symmetric and antisymmetric combinations
\bea
\psi_+&=& d_{-1}^{-A(1)}|0^{-}_{R}\rangle|0^{-}_{R}\rangle + d_{-1}^{-A(2)}|0^{-}_{R}\rangle|0^{-}_{R}\rangle\nn
\psi_-&=& d_{-1}^{-A(1)}|0^{-}_{R}\rangle|0^{-}_{R}\rangle - d_{-1}^{-A(2)}|0^{-}_{R}\rangle|0^{-}_{R}\rangle
\eea
The overall state must be symmetric under the interchange of copies $(1)\leftrightarrow (2)$. Thus if the right sector state is symmetric under this interchange then we must take $\psi_+$ from the above, while if the right sector state is antisymmetric then we must take $\psi_-$. In either case we just get two $j=3/2$ representations, from the two values of the index $A$. This corresponds to the term $2\chi^{L}_{j=3/2}(y)$ in the RHS of (\ref{11 sector chara}). 

Proceeding in this way, we find for the left sector the lowest weight states
(\ref{lowest states L})
\bea\label{left lowest}
2\chi^{L}_{j=3/2} &:&
d_{-1}^{-A(1)}|0^{-}_{R}\rangle|0^{-}_{R}\rangle\nn
4\chi^{L}_{j=1} &:&
d_{-1}^{-A(1)}(d_0^{+B(1)}-d_0^{+B(2)})|0^{-}_{R}\rangle|0^{-}_{R}\rangle
\eea
where we note that the symmetrization under $(1)\leftrightarrow (2)$ will be done later.

For the right sector, the  lowest weight states are given by the condition $\bar d^{-A}_{0}|\phi\rangle=0$ in (\ref{lowest states L}) 
\bea\label{right lowest}
\bar \chi^{R}_{\bar j_3=-1} &:&
|\bar 0^{-}_{R}\rangle|\bar 0^{-}_{R}\rangle\nn
2\bar \chi^{R}_{\bar j_{3}=-1/2} &:&
(\bar d_0^{+A(1)}-\bar d_0^{+A(2)})|\bar 0^{-}_{R}\rangle|\bar 0^{-}_{R}\rangle\nn
\bar \chi^{R}_{\bar j_{3}=0} &:& (\bar d_0^{++(1)}-\bar d_0^{++(2)})(\bar d_0^{+-(1)}-\bar d_0^{+-(2)})|\bar 0^{-}_{R}\rangle|\bar 0^{-}_{R}\rangle
\eea
For this sector the lowest weight states automatically have a definite symmetry under $(1)\leftrightarrow (2)$.

Finally, as noted above we get the correctly symmetrized states of the orbifiold theory by taking an appropriately symmetrized representation from the left and multiplying it with a representation from the right, so that the overall state is symmetric under $(1)\leftrightarrow (2)$.  In this way, we get the complete subspace of the lowest weight states of the global zero mode subalgebra defined by the operators (\ref{global modes for chara}).

\subsection{Global zero mode multiplets for one doubly wound copy at level one}\label{N=(2) multiplets}

In this subsection, we repeat the above computation for the twist sector where we have one doubly wound copy of the CFT; i.e., the sector $N=(2)$. 

This time in the expression (\ref{partition function}) we keep only the contribution of terms with  $n=2$ and collect the terms with power $p^2$. This gives the contribution from the winding sector  $N=(2)$. Expressing the result in terms of characters, we find
\bea\label{2 sector chara}
Z(N=(2);h=1,\bar h=0)&=&(y-8+y^{-1})(y^{1/2}-y^{-1/2})^4(\bar y^{1/2}-\bar y^{-1/2})^2\nn
&=&\left(\chi^{L}_{j=3/2}(y)+8\chi^{L}_{j=1}(y)\right)\bar \chi^{R}_{\bar j_{3}=-1/2}(\bar y)
\eea
The  lowest weight states on the left satisfying (\ref{lowest states L}) are
\bea\label{left lowest doubly}
\chi^{L}_{j=3/2} &:&
d_{-1/2}^{-+}d_{-1/2}^{--}|0^{2-}_{R}\rangle\nn
8\chi^{L}_{j=1} &:&
d_{-1}^{-A}|0^{2-}_{R}\rangle\nn
&& 
d_{-1/2}^{-+}\alpha_{- \dot A,-1/2}|0^{2-}_{R}\rangle\nn
&& 
d_{-1/2}^{--}\alpha_{+ \dot A,-1/2}|0^{2-}_{R}\rangle\nn
&& 
\left(d_{-1/2}^{--}\alpha_{- \dot A,-1/2}-d_{-1/2}^{-+}\alpha_{+ \dot A,-1/2}\right)|0^{2-}_{R}\rangle
\eea
The lowest weight states on the right are
\be\label{right lowest doubly}
\bar \chi^{R}_{\bar j_{3}=-1/2} :
|\bar 0^{2-}_{R}\rangle
\ee
This time there is just one doubly wound copy of the CFT, so we do not have to symmetrize or antisymmetrize the left and right sectors. Any of the above above left states can by tensored any of the above right states to give a lowest weight states of the global zero mode subalgebra defined by the operators (\ref{global modes for chara}).

\section{The effect of twist operator}\label{s twist}

The deformation operator (\ref{D 1/4}) contains a twist $\sigma_2$. The action of this twist was studied in \cite{Avery:2010er,Avery:2010hs}. Here we recall some results about this action which will be of use to us later in the computation of $E^{(2)}$.

We consider only the left sector. Start in the twist sector $N=(1,1)$ where we have two singly wound copies of the CFT. Let the initial state be the Ramond ground state $|0^{-}_{R}\rangle|0^{-}_{R}\rangle$. Let us apply the twist operator $\sigma^+_2$ at the position $w_0$ on the cylinder. This action generates the state $|\chi\rangle$ whose formal structure is as follows
\bea\label{chi}
|\chi\rangle=\sigma^{+}_{2}(w_{0})|0^{-}_{R}\rangle|0^{-}_{R}\rangle=|0^{2-}_{R}\rangle+ a O_{-1/2}|0^{2-}_{R}\rangle +  a^2 O_{-1}|0^{2-}_{R}\rangle+\ldots
\eea
where  $a=e^{w_{0}/2}$ and  the  $O_{-k}$ are operators of dimension $k$. The full expression for $|\chi\rangle$ was found in closed form in  \cite{Avery:2010er,Avery:2010hs}.

We can also start with an initial state which contains one oscillator excitation on the vacuum $|0^{-}_{R}\rangle|0^{-}_{R}\rangle$. For a bosonic oscillator with mode number $n<0$ which is placed on copy $1$ we find
\bea\label{a copy 1}
&&\sigma^{+}_{2}(w_{0})\, \alpha^{(1)}_{A\dot A,n}\, |0^{-}_{R}\rangle|0^{-}_{R}\rangle\nn
&=&\Big(\frac{1}{2}\alpha_{A\dot A,n}+\sum_{p'\leq -1}\frac{i}{\pi}\frac{\Gamma[\frac{1}{2}-n]}{\Gamma[-n]}
\frac{\Gamma[-\frac{1}{2}-p']}{\Gamma[-p']}\frac{a^{2(n-p')-1}}{2n-2p'-1}
\alpha_{A\dot A,p'+1/2}\Big)|\chi\rangle\nn
&\equiv &\sum_{p}f^{B}_{a}[n,p]\, \alpha_{A\dot A,p/2}\, |\chi\rangle
\eea
where $f^{B}_{a}[n,p]$ is defined by the coefficients in the second line.
Here the $\alpha_{A\dot A,p/2}$ are bosonic oscillators acting on the twist sector $N=(2)$ where we have one doubly wound copy of the CFT. Note that we restrict the initial oscillator mode to $n<0$ since we assume that the zero mode of the bosonic oscillators annihilate the vacuum; this corresponds to working in the sector where our states have no $U(1)$ charges coming from the translational symmetries along the four directions of the $T^4$. 

If the initial oscillator is placed on copy $2$ instead of copy $1$, we get a similar result but with the replacement  $a\r -a$ in (\ref{a copy 1})
\be\label{a copy 2}
\sigma^{+}_{2}(w_{0})\, \alpha^{(2)}_{A\dot A,n}\, |0^{-}_{R}\rangle|0^{-}_{R}\rangle=\sum_{p}f^{B}_{-a}[n,p]\, \alpha_{A\dot A,p/2}\, |\chi\rangle
\ee

We can start with one fermionic oscillator instead of the bosonic oscillator. 
For the fermionic excitation $d^{+A}_{n}$ on copy 1 with $n\leq 0$ we get
\bea\label{d+ copy 1}
&&\sigma^{+}_{2}(w_{0})d^{+A(1)}_{n}|0^{-}_{R}\rangle|0^{-}_{R}\rangle\nn
&=&\Big(\frac{1}{2}d^{+A}_{n}+\sum_{p'\leq -1}\frac{i}{\pi}\frac{\Gamma[\frac{1}{2}-n]}{\Gamma[1-n]}
\frac{\Gamma[\frac{1}{2}-p']}{\Gamma[-p']}\frac{a^{2(n-p')-1}}{2n-2p'-1}
d^{+A}_{p'+1/2}\Big)|\chi\rangle\nn
&=&\sum_{p}f^{F+}_{a}[n,p]d^{+A}_{p/2}|\chi\rangle
\eea
If this initial excitation is placed on copy 2, we replace $a$ by $-a$
\be\label{d+ copy 2}
\sigma^{+}_{2}(w_{0})d^{+A(2)}_{n}|0^{-}_{R}\rangle|0^{-}_{R}\rangle
=\sum_{p}f^{F+}_{-a}[n,p]d^{+A}_{p/2}|\chi\rangle
\ee
For states with one fermionic excitation $d^{-A}_{n}$ on copy 1 with $n\leq 0$
\bea\label{d- copy 1}
&&\sigma^{+}_{2}(w_{0})d^{-A(1)}_{n}|0^{-}_{R}\rangle|0^{-}_{R}\rangle\nn
&=&\Big(\frac{1}{2}d^{-A}_{n}+\sum_{p'\leq -1}\frac{i}{\pi}\frac{\Gamma[\frac{1}{2}-n]}{\Gamma[-n]}
\frac{\Gamma[-\frac{1}{2}-p']}{\Gamma[-p']}\frac{a^{2(n-p')-1}}{2n-2p'-1}
d^{-A}_{p'+1/2}\Big)|\chi\rangle\nn
&=&\sum_{p}f^{F-}_{a}[n,p]d^{-A}_{p/2}|\chi\rangle
\eea
For this excitation on copy 2, we replace $a$ by $-a$
\be\label{d- copy 2}
\sigma^{+}_{2}(w_{0})d^{-A(2)}_{n}|0^{-}_{R}\rangle|0^{-}_{R}\rangle
=\sum_{p}f^{F-}_{-a}[n,p]d^{-A}_{p/2}|\chi\rangle
\ee

We can also start with initial states that have more than one oscillator excitation; i.e., we can compute
\be
\sigma^{+}_{2}(w_{0})\prod_{i}O^{(c_i)}_{i,-n_{i}}|0^{-}_{R}\rangle|0^{-}_{R}\rangle, ~~~c_i=1,2
\ee
where $c_i$ denoted the copy in which the oscillator $O_{i, -n_i}$ is  placed. The general method to compute the final state in this situation was given in \cite{Avery:2010hs}. There are terms where each oscillator in the initial state is moved separately to the final state as indicated in the relations (\ref{a copy 1}-\ref{d- copy 2}) discussed above where we had only one initial operator. There are additional terms that arise from the contractions between oscillator modes in the initial state. It will turn out that for the cases we encounter below, there are no contractions between the modes in the initial state. We will need a subset of the coefficients $f$ in the relations (\ref{a copy 1}-\ref{d- copy 2}). We write them in the following form which will be of use to us:
\be\label{f- -1}
\sum_{p}f^{F-}_{a}[-1,p]d_{p/2}=-\frac{i}{2}a^{-1}d_{-1/2}+\frac{1}{2}d_{-1}+O(a)
\ee
\be\label{f- 0}
\sum_{p}f^{F-}_{a}[0,p]d_{p/2}=\frac{1}{2}d_{0}
\ee
\be\label{f+ -1}
\sum_{p}f^{F+}_{a}[-1,p]d_{p/2}=-\frac{i}{4}a^{-1}d_{-1/2}+\frac{1}{2}d_{-1}+O(a)
\ee
\be\label{f+ 0}
\sum_{p}f^{F+}_{a}[0,p]d_{p/2}=\frac{1}{2}d_{0}+\frac{i}{2}a d_{-1/2}+O(a^3)
\ee

The dependence of the above expressions on the parameter $a$ is rather simple; the power of $a$ just keeps track of the dimension of the operator it multiplies. We will find two uses of this parameter. First, by taking the limit $a\rightarrow 0$ we can  project onto states of a  certain dimension. Second,   the replacement $a\rightarrow -a$, interchanges an excitation between copies 1 and  2; this will help us in making excitations that are symmetric or antisymmetric between the two copies.  

\section{The lifting and the long multiplet at level 1}\label{section lifting}

With the tools we have collected, we can now move on to the computation of lifting for the level $1$ states.   In section \ref{s right mover}, we discuss the  part of the amplitudes involving the right movers. In section \ref{s left mover}, we will consider the left movers. In section \ref{long multiplet}, we will combine these results to get the structure of the lifted multiplet and the value of the  lifting.

\subsection{The right sector of states having nozero lifting}\label{s right mover}

The  level $1$ states we are studying have conformal dimensions $(h, \bar h)=({c\over 24}+1, {c\over 24})$. Thus the right movers are in the Ramond ground state. In section \ref{s global multiplet}, we had organized all the right moving ground states in terms of multiplets created by the global modes $\bar d^{\bar \alpha A}_0$. For the winding sector $N=(1,1)$, the lowest weight states of these multiplets were given in (\ref{right lowest}):
\bea\label{right lowest 1}
|\bar 0^{-}_{R}\rangle|\bar 0^{-}_{R}\rangle\nn
(\bar d_0^{+A(1)}-\bar d_0^{+A(2)})|\bar 0^{-}_{R}\rangle|\bar 0^{-}_{R}\rangle\nn
(\bar d_0^{++(1)}-\bar d_0^{++(2)})(\bar d_0^{+-(1)}-\bar d_0^{+-(2)})|\bar 0^{-}_{R}\rangle|\bar 0^{-}_{R}\rangle
\eea
For the winding sector $N=(2)$ the lowest weight state was given in  (\ref{right lowest doubly}):
\bea\label{right lowest doubly 1}
|\bar 0^{2-}_{R}\rangle
\eea
The right mover of the operator (\ref{GN p s}) $\bar G^{\bar \alpha (P)}_{\dot A,0}$ is
\be
\mathcal P \bar \sigma^{\bar \alpha}
\ee
where we drop the second projection operator since it is acting on states with dimension (\ref{sthree}).
Therefore in the long multiplet structure (\ref{multiplets diagram}), the two `raising' operators have the same operator $\mathcal P\bar \sigma^{+}$ in the right sector. The two `lowering' operators $\bar G^{-}_{\pm}$ have the same operator $\mathcal P\bar \sigma^{-}$ in the right sector. Thus the long multiplet structure (\ref{sthree}) for the right movers is as given in the following diagram:

\be\label{multiplets diagram right}
\begin{tikzcd}
                                                & \phi^R_{+} \arrow[dr,"\mathcal P\bar \sigma^{+}"]\\
\phi^R \arrow[ur, "\mathcal P\bar \sigma^{+}"]  \arrow{dr}[swap]{\mathcal P\bar \sigma^{+}}& &  \phi^R_{+-}
\\
                                                & \phi^R_{-} \arrow{ur}[swap]{\mathcal P\bar \sigma^{+}}
\end{tikzcd}
~~~~~~~~~
\begin{tikzcd}
                                                & \arrow{dl} [swap]{\mathcal P\bar \sigma^{-}} \phi^R_{+} \\
\phi^R   & &  \arrow{ul}[swap]{\mathcal P\bar \sigma^{-}} \phi^R_{+-} \arrow{dl}{\mathcal P\bar \sigma^{-}}
\\
                                                & 
                    \arrow{ul}{\mathcal P\bar \sigma^{-}} \phi^R_{-} 
\end{tikzcd}
\ee

In the appendix \ref{app prop r}, we will find the following results:

\b

(i) The right moving sector of the states in the long multiplet in fig.\ref{multiplets diagram right} are
\bea\label{right mover}
|\phi^{R}\rangle&=&|\bar 0^{-}_{R}\rangle|\bar 0^{-}_{R}\rangle\nn
|\phi^{R}_{+}\rangle=|\phi^{R}_{-}\rangle&=&|\bar 0^{2-}_{R}\rangle\nn
|\phi^{R}_{+-}\rangle&=&\frac{1}{2}(\bar d_0^{++(1)}-\bar d_0^{++(2)})(\bar d_0^{+-(1)}-\bar d_0^{+-(2)})|\bar 0^{-}_{R}\rangle|\bar 0^{-}_{R}\rangle
\eea

\b

(ii) States which have the right moving sector
\be\label{r mover zero lifting}
(\bar d_0^{+A(1)}-\bar d_0^{+A(2)})|\bar 0^{-}_{R}\rangle|\bar 0^{-}_{R}\rangle
\ee
have zero lift.

\b

Let us summarize the structure we have found in Appendix \ref{app prop r} for the right sector of the long multiplet. Consider  figure (\ref{multiplets diagram right}), which shows the right moving part of the long multiplet structure given in figure  (\ref{multiplets diagram}); these figures show how  four short multiplets join into a long multiplet. We start with the state $ |\phi^{R}\rangle=|\bar 0^{-}_{R}\rangle|\bar 0^{-}_{R}\rangle$ at the bottom of the multiplet. From (\ref{anni bottom}) we see that this state is annihilated by $\mathcal P\bar \sigma^{-}$
\be\label{right bottom anni}
\mathcal P\bar \sigma^{-} |\phi^{R}\rangle=0
\ee
so we cannot move along the multiplet by applying the $\bar G^{-}_{\pm}$. We can apply the $\bar G^{+}_{\pm}$ to $ |\phi^{R}\rangle$, and  we find from (\ref{p move bottom}) that this will take us to a state where the right moving part is $|\bar 0^{2-}_{R}\rangle$. Since the two raising operators $\bar G^{+}_{\pm}$ have the same action on the right side, the two states $\phi_{+}$ and $\phi_{-}$ are the same on the right, and will differ only in their left part.
Finally, we see that a second application of the $\bar G^{+}_{\pm}$ will give the right moving state $\phi^{R}_{+-}$.

From (\ref{top right anni}) we see that the state $\phi^{R}_{+-}$ is annihilated by $\mathcal P\bar \sigma^{+}$
\be\label{right top anni}
\mathcal P\bar \sigma^{+} |\phi_{+-}^{R}\rangle=0
\ee
so we cannot move further along the multiplet by applying the $\bar G^{+}_{\pm}$.

We observe that in the sector $N=(1,1)$ where we have two singly wound copies of the CFT, the right moving ground states (\ref{right mover}) are symmetric between the two copies. Since the overall state must be symmetric,  the left moving sector of these long multiplet states must be symmetric as well.

\subsection{The left sector of states with nozero lift}\label{s left mover}

We now consider the left sector. In section\,\ref{s global multiplet}, we had organized the states into representations generated by the global modes $d_{0}^{\alpha A},J^{a}_{0},G^{\alpha}_{\dot A,0}$. The lowest weight states in the winding sector $N=(1,1)$ were found in (\ref{left lowest}):
\bea\label{left lowest 1}
d_{-1}^{-A(1)}|0^{-}_{R}\rangle|0^{-}_{R}\rangle\nn
d_{-1}^{-A(1)}(d_0^{+B(1)}-d_0^{+B(2)})|0^{-}_{R}\rangle|0^{-}_{R}\rangle
\eea
where it is understood that these states have to be appropriately symmetrized/antisymmetrized between the two copies. In the previous subsection, we had noted that for states in a long multiplet the state on the right was symmetric between the two copies. The overall orbifold symmetry then says that the left sector must be symmetric as well. Thus if we are looking for long multiplets we must symmetrize the states  in (\ref{left lowest 1}). After this symmetrzation we find that the states in the first line of (\ref{left lowest 1}) become global states. In  Appendix \ref{app global unlifted} we show that these global states are unlifted.   The states in the second line can be decomposed into a  triplet and singlet of the $SU(2)$ charge $A$. The lowest weight state of the triplet is 
\be
(d_{-1}^{--(1)}-d_{-1}^{--(2)})(d_0^{+-(1)}-d_0^{+-(2)})|0^{-}_{R}\rangle|0^{-}_{R}\rangle\label{triplet}
\ee
and for the singlet is
\be
\left[(d_{-1}^{-+(1)}-d_{-1}^{-+(2)})(d_0^{+-(1)}-d_0^{+-(2)})-(d_{-1}^{--(1)}-d_{-1}^{--(2)})(d_0^{++(1)}-d_0^{++(2)})\right]|0^{-}_{R}\rangle|0^{-}_{R}\rangle\label{singlet}
\ee
The singlet has zero lift because it is given by global mode excitations on a Ramond ground state
\bea\label{singlet is global}
&&\left[(d_{-1}^{-+(1)}-d_{-1}^{-+(2)})(d_0^{+-(1)}-d_0^{+-(2)})-(d_{-1}^{--(1)}-d_{-1}^{--(2)})(d_0^{++(1)}-d_0^{++(2)})\right]|0^{-}_{R}\rangle|0^{-}_{R}\rangle\nn
&=& J_{-1}^-(d_{0}^{++(1)}-d_{0}^{++(2)})(d_0^{+-(1)}-d_0^{+-(2)})|0^{-}_{R}\rangle|0^{-}_{R}\rangle
\eea
Thus the left sector of the states with nonzero lift must lie in the space of the triplet $A$ charge,  given in (\ref{triplet}). Since all members of the triplet must have the same lift ($A$ charge automorphism), we need to calculate the lift for only one of these states. 

The lifting matrix (\ref{liftmatrix 1})  involves the operators $\bar G^{\bar \alpha (P)}_{\dot A,0}$ (\ref{GN p s}).
Consider the left sector of these operators. 
\be
\pi \mathcal P G^{+}_{\dot A,-\frac{1}{2}}\sigma^{-}
\ee
We need to compute the action of the above operator on the left sector state whose lift we are computing. For the triplet (\ref{triplet}), we show in appendix \ref{app prop l} that:
\bea\label{left amplitude}
\mathcal P G^{+}_{\dot A,-\frac{1}{2}}\sigma^{-}(d_{-1}^{--(1)}-d_{-1}^{--(2)})(d_0^{+-(1)}-d_0^{+-(2)})|0^{-}_{R}\rangle|0^{-}_{R}\rangle=-i d^{--}_{-1/2}\alpha_{+\dot A,-1/2}|0^{2-}_{R}\rangle
\eea

Based on (\ref{left amplitude}), define the following normalized left moving states
\bea\label{left movers}
|\phi^{L}\rangle=|\phi^{L}_{+-}\rangle
&=&\frac{1}{2}(d_{-1}^{--(1)}-d_{-1}^{--(2)})(d_0^{+-(1)}-d_0^{+-(2)})|0^{-}_{R}\rangle|0^{-}_{R}\rangle\nn
|\phi^{L}_{+}\rangle&=&\frac{1}{\sqrt{2}} d^{--}_{-1/2}\alpha_{++,-1/2}|0^{2-}_{R}\rangle\nn
|\phi^{L}_{-}\rangle&=&\frac{1}{\sqrt{2}} d^{--}_{-1/2}\alpha_{+-,-1/2}|0^{2-}_{R}\rangle
\eea
Thus eq.\,(\ref{left amplitude}) is 
\be\label{left amplitude 2}
\pi\mathcal P G^{+}_{\dot A,-\frac{1}{2}}\sigma^{-}|\phi^{L}\rangle=-\frac{i\pi}{\sqrt{2}}|\phi^{L}_{\dot A}\rangle;
~~~~~~~~~~~
\pi\mathcal P G^{+}_{\dot A,-\frac{1}{2}}\sigma^{-}|\phi^{L}_{+-}\rangle=-\frac{i\pi}{\sqrt{2}}|\phi^{L}_{\dot A}\rangle
\ee

The properties  in (\ref{left amplitude 2}) can be summarized into the following diagram
\be\label{multiplets diagram left}
\begin{tikzcd}
                                                & \phi^L_{+} \\
\phi^L \arrow[ur, "\pi\mathcal P G^{+}_{+,-1/2}\sigma^{-}"]  \arrow{dr}[swap]{\pi\mathcal P G^{+}_{-,-1/2}\sigma^{-}}& &  \arrow{ul}[swap]{\pi\mathcal P G^{+}_{+,-1/2}\sigma^{-}} \phi^L_{+-} \arrow{dl}{\pi\mathcal P G^{+}_{-,-1/2}\sigma^{-}}
\\
                                                & \phi_{-}^L 
\end{tikzcd}
\ee

\subsection{The long multiplet and its lifting}\label{long multiplet}

We have seen that the long multiplets at level $1$ form a triplet of $A$ charge. Thus we have to compute just one independent lift.  To find the lift we must  combine the left and right sectors.

In section \ref{s right mover}, we found that  the right movers of the long multiplet were given by (\ref{right mover})
and in section \ref{s left mover} we defined the left parts of a set of  states in  (\ref{left movers}).
We will now combine the right and left parts of these states in an appropriate fashion. The long multiplet contains four short multiplets. The lowest weight states of these four short multiplets will be 
\bea
|\phi\rangle&=&\frac{1}{2}(d_{-1}^{--(1)}-d_{-1}^{--(2)})(d_0^{+-(1)}-d_0^{+-(2)})|0^{-}_{R}\rangle|0^{-}_{R}\rangle|\bar 0^{-}_{R}\rangle|\bar 0^{-}_{R}\rangle\nn
|\phi_{+}\rangle&=&\frac{1}{\sqrt{2}} d^{--}_{-1/2}\alpha_{++,-1/2}|0^{2-}_{R}\rangle|\bar 0^{2-}_{R}\rangle\nn
|\phi_{-}\rangle&=&\frac{1}{\sqrt{2}} d^{--}_{-1/2}\alpha_{+-,-1/2}|0^{2-}_{R}\rangle|\bar 0^{2-}_{R}\rangle\nn
|\phi_{+-}\rangle&=&\nn
\frac{1}{4}&&\hskip-30pt (d_{-1}^{--(1)}-d_{-1}^{--(2)})(d_0^{+-(1)}-d_0^{+-(2)})(\bar d_0^{++(1)}-\bar d_0^{++(2)})(\bar d_0^{+-(1)}-\bar d_0^{+-(2)})|0^{-}_{R}\rangle|0^{-}_{R}\rangle|\bar 0^{-}_{R}\rangle|\bar 0^{-}_{R}\rangle\nn
\label{fourstates}
\eea

From (\ref{multiplets diagram right}) and (\ref{multiplets diagram left}), one can see that the above states satisfy the following relations
\be\label{multiplets diagram full}
\begin{tikzcd}
                                                & \phi_{+} \\
\phi \arrow[ur, "\pi\mathcal P G^{+}_{+,-1/2}\sigma^{-}\bar \sigma^{+}"]  \arrow{dr}[swap]{\pi\mathcal P G^{+}_{-,-1/2}\sigma^{-}\bar \sigma^{+}}& &  \arrow{ul}[swap]{\pi\mathcal P G^{+}_{+,-1/2}\sigma^{-}\bar \sigma^{-}} \phi_{+-} \arrow{dl}{\pi\mathcal P G^{+}_{-,-1/2}\sigma^{-}\bar \sigma^{-}}
\\
                                                & \phi_{-}
\end{tikzcd}
\ee
Now we will prove that the four states (\ref{fourstates}) indeed form a long multiplet. Let's first show that $\phi$ is the bottom member of the long multiplet. We had noted in (\ref{right mover}) that  the right mover of the bottom member of a long multiplet must be  $\phi^R$. Now let's look for possible left movers.
From section \ref{s left mover}, we know the only possible left movers for long multiplets are the triplets with lowest weight state (\ref{triplet}). Combining the three states from the triplets with the right mover $\phi^R$ gives us three possible states as the bottom members of long multiplets.  We will thus get a triplet of long multiplets. The state $\phi$ listed above is the member of this triplet with lowest $A$ charge ($A=-1$).

In a similar way we can show that $\phi_{+-}$ is the top member of a long multiplet. First we note that the right mover of the top member of a long multiplet must be  $\phi^R_{+-}$ as in (\ref{right mover}). The only possible left movers are the triplets with lowest weight state (\ref{triplet}). Thus combining the left mover (\ref{triplet}) and the right mover $\phi^R_{+-}$ gives the top member $\phi_{+-}$ of a long multiplet.

To find the middle member $\phi_{+}$ of the multiplet, we can apply $\bar G^{+(P)}_+=\pi\mathcal P G^{+}_{+,-1/2}\sigma^{-}\bar \sigma^{+}$ to the bottom member $\phi$ or apply $\bar G^{-(P)}_+=\pi\mathcal P G^{+}_{+,-1/2}\sigma^{-}\bar \sigma^{-}$ to the top member $\phi_{+-}$. These two ways are verified in relations (\ref{multiplets diagram full}). Thus  $\phi_{+}$ in (\ref{fourstates}) is indeed a middle member of the long multiplet. Similarly, the two ways to find the middle member $\phi_{-}$ are verified in (\ref{fourstates}). Thus $\phi_{-}$ in (\ref{fourstates}) is a correct middle member of the long multiplet.

Since all the members in a long multiplet must be  eigenstates of the lifting matrix, we can use the relation  (\ref{lifting norm}) to get the lifting. 
All the members have the same lift, so we only need to calculate the lift for the bottom member $\phi$. Using the first expression in (\ref{lifting norm}), we have
\bea\label{lifting for long multi}
E^{(2)}&=&2 \lambda^2   
\Big(\Big|\bar G^{+(P)}_{+,0} |\phi\rangle\Big|^2+\Big|\bar G^{-(P)}_{-,0} |\phi\rangle\Big|^2\Big)\nn
&=&2 \lambda^2   
\Big(\Big|\pi\mathcal P G^{+}_{+,-1/2}\sigma^{-}\bar \sigma^{+} |\phi\rangle\Big|^2+\Big|\pi\mathcal P G^{+}_{-,-1/2}\sigma^{-}\bar \sigma^{-} |\phi\rangle\Big|^2\Big)
\eea
For the first term, from (\ref{p move bottom}) and (\ref{left amplitude 2}), we have
\be
\pi\mathcal P G^{+}_{+,-\frac{1}{2}}\sigma^{-}\bar \sigma^{+}|\phi\rangle=-\frac{i\pi}{\sqrt{2}}|\phi_{+}\rangle
\ee
The second term in the final expression in (\ref{lifting for long multi}) is zero due to (\ref{right bottom anni}). Thus the lift is
\be\label{lifting final}
E^{(2)}= \lambda^2  \pi^2
\ee

To summarize,  we find three long multiplets, which form a triplet of charge $A$. The member of this triplet  with lowest $A$ charge  ($A=-1$) is given in  (\ref{fourstates}). All the states joining into long multiplets have the same lift (\ref{lifting final}).

\section{Unlifted states and the index}

We have studied the states at level 1 and found states that are lifted and states that remain unlifted. In this section we will count the number of states of each kind. We will also examine what we can learn from an index type computation which provides a lower bound on the number of unlifted states. We will find that at level 1 the lower bound is actually saturated, and all these unlifted states are in fact global modes.

In section \ref{section n of states}, we will count the total number of states at level 1. In section \ref{section global states}, we will find all the global states. In section \ref{section bound}, we will find a lower bound of the unlifted states from an index computation. In section \ref{section unlifted orbifold}, we will find that the number of unlifted states from our perturbation calculation saturates that lower bound.

\subsection{The number of states}\label{section n of states}
In this subsection, we will count the number of states at level 1. One can see that if we take $y=\bar y=-1$ in the character (\ref{global character}), we  get the number of states. 
For the $N=(1,1)$ sector (\ref{11 sector chara}), the number of states at level 1 is
\be\label{11 states}
\left(2\chi^{L}_{j=3/2}(y)+4\chi^{L}_{j=1}(y)\right)\left(\bar \chi^{R}_{\bar j_{3}=-1}(\bar y)+2\bar \chi^{R}_{\bar j_{3}=-1/2}(\bar y)+\bar \chi^{R}_{\bar j_{3}=0}(\bar y)\right)\Big|_{y=\bar y=-1}
=2048
\ee
For the $N=(2)$ sector (\ref{2 sector chara}), the number of states at level 1 is
\be\label{2 states}
\left(\chi^{L}_{j=3/2}(y)+8\chi^{L}_{j=1}(y)\right)\bar \chi^{R}_{\bar j_{3}=-1/2}(\bar y)\Big|_{y=\bar y=-1}
=640
\ee
Thus the total number of states is
\be\label{number of states}
\text{Number of states} = 2688
\ee

\subsection{The number of global states}\label{section global states}
In this subsection, we will count the number of global states.\footnote{We thank Nathan Benjamin and Xinan Zhou for useful discussions on the characters created by global modes.} 
These states are unlifted states due to the contracted large $\mathcal N=4$ superconformal symmetry.

The first class of unlifted states is in the $(1,1)$ sector
\be
2\chi^{L}_{j=3/2}(y)(\bar \chi^{R}_{\bar j_{3}=-1}(\bar y)+\bar \chi^{R}_{\bar j_{3}=0}(\bar y))
\ee
where the states can be read from (\ref{left lowest}) and (\ref{right lowest}). In this class, the global excitation $d_{-1}^{-A}$ acts on the Ramond ground states. Taking $y=\bar y =-1$ gives the number of states in the class, which is $512$. 

The second class is in the $(1,1)$ sector
\be
(1,1):~2 \chi^{L}_{j=3/2}(y)2 \bar \chi^{R}_{\bar j_{3}=-1/2}(\bar y)
\ee
The states are
\bea
&&(d^{-A(1)}_{-1}-d^{-A(2)}_{-1})(\bar d^{+B(1)}_{0}-\bar d^{+B(2)}_{0})|0^-_{R}\rangle| 0^-_{R}\rangle|\bar 0^-_{R}\rangle|\bar 0^-_{R}\rangle\nn
&=&J^{-}_{-1}(d^{+A(1)}_{0}-d^{+A(2)}_{0})(\bar d^{+B(1)}_{0}-\bar d^{+B(2)}_{0})|0^-_{R}\rangle| 0^-_{R}\rangle|\bar 0^-_{R}\rangle|\bar 0^-_{R}\rangle
\eea
which can be written as a global excitation $J^{-}_{-1}$ acting on the Ramond ground state. The number of states in the this class in $512$.

The third class is in the $(1,1)$ sector
\be
4\chi^{L}_{j=1}(y)2\bar \chi^{R}_{\bar j_{3}=-1/2}(\bar y)
\ee
where the states can be read from (\ref{left lowest}) and (\ref{right lowest}). In this class, the global excitation $d_{-1}^{-A}$ acts on the Ramond ground states. The number of states is $512$.

The fourth class is in the $(1,1)$ sector
\be
\chi^{L}_{j=1}(y)(\bar \chi^{R}_{\bar j_{3}=-1}(\bar y)+\bar \chi^{R}_{\bar j_{3}=0}(\bar y))
\ee
where the left mover is given by (\ref{singlet is global}) and right mover is given by (\ref{right lowest}). As we can see from (\ref{singlet is global}), these are global states. The number of states is $128$. 

The fifth class is in the $(2)$ sector
\be
\chi^{L}_{j=3/2}(y)\bar \chi^{R}_{\bar j_{3}=-1/2}(\bar y)
\ee
where the state can be read from (\ref{left lowest doubly}) and (\ref{right lowest doubly}); the state  is $d_{-1/2}^{-+}d_{-1/2}^{--}|0^{2-}_{R}\rangle|\bar 0^{2-}_{R}\rangle=-2J^-_{-1}|0^{2-}_{R}\rangle|\bar 0^{2-}_{R}\rangle$. It is a global state. The number of states in this class is $128$.

The last class is in the $(2)$ sector
\be
2\chi^{L}_{j=1}(y)\bar \chi^{R}_{\bar j_{3}=-1/2}(\bar y)
\ee
where state can be read from (\ref{left lowest doubly}) and (\ref{right lowest doubly}); it   is $d_{-1}^{-A}|0^{2-}_{R}\rangle|\bar 0^{2-}_{R}\rangle$. It is a global state. The number of states in this class is $128$. 

Thus the total number of global states from the above six classes is
\be
\text{Number of global states} =512+512+512+128+128+128=1920
\ee

\subsection{A lower bound on the  number of unlifted states}\label{section bound}

We have already noted that global states will remain unlifted. A lower bound on the number of unlifted states can be found from an index \cite{sv,mms,b,bh}. The index computation may, however tell us that there are additional (i.e., nonglobal)  that will not lift. We will now see that at level 1, there are no such additional unlifted states  predicted by the index.

To see this, let us recall how the index computation is done. Consider the  exact supercharge operators $\bar G^{+}_{\dot A}$, $\dot A=+,-$ of the perturbed CFT. These operators are the ones that join four short multiplets  into a long multiplet. Let us see the structure of the set of states that will join into a long multiplet.
Each of these operators $\bar G^{+}_{\dot A}$ increases the $SU(2)_R$ charge by 1/2, while it does not change the $SU(2)_L$ charge.
Thus the four short multiplets joining into a long multiplet must have the same left moving character but their right moving characters  will be as follows:
\be\label{right chi long}
\bar \chi^{R}_{\bar j_{3}}~~~~~~2\bar \chi^{R}_{\bar j_{3}+1/2}~~~~~\bar \chi^{R}_{\bar j_{3}+1}
\ee
Whenever we can group states in the manner indicated by such a set of characters, then we find a set of states that have the charges to join into a long multiplet. We therefore exclude such sets of states from the index. If there are states left over that {\it cannot} group into a set with characters (\ref{right chi long}), then we count those states in the index, since they cannot possibly join into a long multiplet and lift.

We will now see that in our case all the states which are not global states actually fall into sets having the characters (\ref{right chi long}), and so they will not be be counted in the index. 
 
The set of all states at level 1 was counted in section \ref{section n of states}. Let us exclude from this set the global states counted in section \ref{section global states}. The remaining states (i.e., nonglobal states) are described by the following characters:
\bea\label{exclude global}
(1,1): && 3\chi^{L}_{j=1}(y)(\bar \chi^{R}_{\bar j_{3}=-1}(\bar y)+\bar \chi^{R}_{\bar j_{3}=0}(\bar y))\nn
(2):&& 6\chi^{L}_{j=1}(y)\bar \chi^{R}_{\bar j_{3}=-1/2}(\bar y)
\eea
 We find that these 12 characters can be grouped into three groups. Each group has four characters whose left movers are $\chi^{L}_{j=1}$ and right movers are (\ref{right chi long}) with $j_{3}=-1$. Thus we see that it is possible to combine all the nonglobal states in (\ref{exclude global}) into three long multiplets. 
 
 To summarize, the index argument does not require any unlifted states at level 1 in addition to the global states. Thus the lower bound on the number of unlifted states provided by the index is the same as the number of global states:
\be\label{lower bound}
\text{Number of unlifted states} \geq  ~\text{Number of global states} =1920
\ee

\subsection{Number of unlifted states at $\lambda^2$}\label{section unlifted orbifold}

Now we note that our computation in section \ref{long multiplet} shows that the states (\ref{exclude global}) which could join into a long multiplet are actually lifted.  
Thus we find that as we perturb away from the orbifold point
\be\label{unlifted orbifold}
\text{Number of unlifted states at order $\lambda^2$} = 1920
\ee
Thus the number of unlifted states saturates the  bound (\ref{lower bound}) found in section \ref{section bound}.
To summarize, for the perturbation calculation at order $\lambda^2$ at level 1, we find that all the unlifted states are  global states and all the remaining states are lifted.

\section{A larger set of deformations}

In all our analysis so far we have consider the deformation of the orbifold CFT by the deformation  (\ref{D 1/4}). This deformation is a singlet under all our global charges. One can consider the more general set of deformations given by the operators
\be\label{D general}
\tilde D=\frac{1}{4}P^{\dot A\dot B}\epsilon_{\alpha\beta}\epsilon_{\bar\alpha \bar\beta} G^{\alpha(0)}_{\dot A, -\h} \bar G^{\bar \alpha(0)}_{\dot B, -\h} \sigma^{\beta \bar\beta}
\ee
The deformation operator must be Hermitian. Some rules for Hermitian conjugation are noted in appendix \ref{app conjugate}. Using these we find  that the matrix $P$ can be parameterized by four real parameters $p_0,p_1,p_2,p_3$
\be\label{param p}
P=p_0 I + \sum_{k=1}^{3} ip_k \sigma_k
\ee
where $\sigma_k$ are the Pauli matrices.
Following the method in \cite{Guo:2019pzk}, one can generalize the lifting matrix (\ref{liftmatrix 1}) for the set of deformations (\ref{D general}):
\bea\label{liftmatrix general}
\tilde E^{(2)}_{ba}&=&P^{\dot A \dot B}P^{\dot C \dot D}\epsilon_{\dot A \dot C}\epsilon_{\dot B \dot D} \lambda^2   
\Big\langle O^{(0)}_{b}\Big|    \Big\{  \bar G^{+(P)\dagger}_{+,0},  \bar G^{+(P)}_{+,0} \Big\} \Big|O^{(0)}_{a}\Big\rangle\nn
&=&P^{\dot A \dot B}P^{\dot C \dot D}\epsilon_{\dot A \dot C}\epsilon_{\dot B \dot D} \lambda^2   
\Big\langle O^{(0)}_{b}\Big|    \Big\{  \bar G^{+(P)\dagger}_{-,0},  \bar G^{+(P)}_{-,0} \Big\} \Big|O^{(0)}_{a}\Big\rangle
\eea
Using the parameters in (\ref{param p}), we have
\be
P^{\dot A \dot B}P^{\dot C \dot D}\epsilon_{\dot A \dot C}\epsilon_{\dot B \dot D}=p_0^2+p_1^2+p_2^2+p_3^2\equiv p^2
\ee
Comparing the lifting matrices (\ref{liftmatrix 1}) and (\ref{liftmatrix general}), one finds that they are proportional to each other:
\be
\tilde E^{(2)}=\frac{p^2}{2}E^{(2)}
\ee
The eigenstates with definite energy and their lift are given by eigenstates and eigenvalues of the lifting matrix. Thus we find that the states which are lifted do not depend on the parameters $p_0, p_i$, and the value of their lift is just given through the invariant $p^2$.  The structure of the long multiplets also does not depend on the choice of $P^{\dot A\dot B}$. 

\section{Discussion}

The D1D5 CFT is a very useful tool in the study of black holes in string theory. However, several aspects of its supersymmetric states remain mysterious. In this paper we have analyzed the supermultiplet structure and lifting of states at level 1,  at second order in the deformation off the orbifold point. We now mention some of the questions about the D1D5 system that we hope to address by our computations and its extensions to higher level states.

\b

{\it The black hole threshold:} The NS sector of the CFT is dual to global $AdS_3\times S^3\times T^4$. At low energies we expect that most states that are extremal at the orbifold point  will be lifted as we perturb away from the orbifold point. This is because in the dual gravity theory there are relatively few supergravity states at low energies, and stringy excitations will be lifted in general. In \cite{gn} the lift of stringy states in the pp-wave approximation was computed in the dual CFT and agreement was found.  

But at energies higher than the black hole threshold we expect a large class of {\it unlifted} states which will account for the entropy of the extremal hole. In the Ramond sector these are states with dimensions $(h, \bar h)=({c\over 24}+n, {c\over 24})$. The index computations of \cite{sv,mms} indeed imply such a large number of unlifted states. 

Now consider the  perturbative computation of the lift, say at second order. There should be some feature of this computation that explains the small number of unlifted states in the NS sector computation mentioned above but leads to  a large number of unlifted states when we consider levels that describe black hole states. It would be very interesting to identify this feature, as it would tell us something about the origin of the large value of black hole entropy. Finding this feature is one of our goals in studying the detailed pattern of lifting.

\b

{\it Twist sectors for lifted and unlifted states:} Consider the index computation of \cite{mms}. If we take the D1, D5, P charges to be coprime, then we see that  the index equals the count of all states in the maximally twisted sector. This does not of course mean that it is the states in the maximally twisted sector that will remain unlifted. In our computation at level 1 we found that the lifted multiplets have half their states in the maximally twisted sector and half in the untwisted sector.

What will be the pattern of lifting in general? For $n_1n_5=N$  large, will the supersymmetric states have support in the highly twisted sectors with twist $\sim N$?  Or will the support be in sectors with intermediate twists $\sim \sqrt{N}$? Or perhaps the support would lie in sectors with low twists $\sim 1$?  The index computation cannot distinguish between these possibilities. Extending our computations of lift to higher levels may shed light on this question. The question is of direct physical relevance to the structure of black holes in the fuzzball paradigm as sectors with low twists correspond to gravity solutions with shallow throats and sectors with high twists correspond to gravity solutions with deep throats. 

More generally we would like to ask: given the set of extremal states at a given level $n$ at the orbifold point, is there an elegant prescription which will tell us which of these states will group together into long multiplets and lift, and what will be the value of the lift?

\b

{\it Separating global modes:} At any point in the moduli space the CFT has a set of raising operators given by elements of the chiral algebra; we have called these `global modes' since they act on all copies of the $c=6$ CFT rather than on just one individual copy. The `global states' obtained by applying these global excitations on Ramond ground states will always remain unlifted. In \cite{Mathur:2011gz} it was shown that the action of global modes is mapped, in the dual gravity theory, to excitations at the `neck' region of the geometry where the outer boundary of the AdS joins flat spacetime. Thus global modes therefore do not modify the structure of a black hole microstate whose nontrivial fuzzball structure is at the bottom of a deep throat. 

In an index computation, one attempts to group states with character sets of the type (\ref{right chi long}) in order to identify how many sets of states can join into long multiplets and lift. Some of the states that might appear to join into long multiplets in this analysis might actually be global states. Such states will not lift, but this fact will not be obvious by looking at the characters representing these states. Thus it is useful to first separate out the global states from the set of all states, and then to use the character grouping (\ref{right chi long}) to develop an index count. We have indeed separated global states first in our analysis of level 1 states. A general analysis of the index after separating global states has been carried out in \cite{b,bh}.  

More generally, it would be useful to be able to identify all states that are global descendants of states at lower levels. In other words, we would like to have a simple prescription for identifying all the primaries of the large ${\mathcal N}=4$ superalgebra; the analysis of lifting can then be confined to these primaries. 

\b

We hope to return to these issues elsewhere, when we consider the lifting at states at higher levels and with larger values of $n_1n_5=N$.

 \section*{Acknowledgements}

We would like to thank Nathan Benjamin, Stefano Giusto, Shaun Hampton, Rodolfo Russo, Alfred Shapere, Ida Zadeh and  Xinan Zhou, for many helpful discussions.  This work is supported in part by DOE grant de-sc0011726.

\appendix

\section{Contracted large $\mathcal N=4$ superconformal algebra}\label{commutators}
We follow the notation in the appendix of \cite{hmz}. The indices $\alpha=(+,-)$ and $\bar \alpha=(+,-)$ correspond to the subgroups $SU(2)_L$ and $SU(2)_R$ arising from rotations on $S^3$. The indices $ A=(+,-)$ and $\dot A=(+,-)$ correspond to the subgroups $SU(2)_1$ and $SU(2)_2$ arising  from rotations in $T^4$. We use the convention
\be
\epsilon_{+-}=1, ~~~\epsilon^{+-}=-1
\ee
The commutation relations for the contracted large $\mathcal N=4$ superconformal algebra are
\bea\label{app com a d}
[\alpha_{A\dot{A},m},\alpha_{B\dot{B},n}] &=& -\frac{c}{6}m\epsilon_{A B}\epsilon_{\dot A \dot{B}}\delta_{m+n,0}\cr
\{d^{\alpha A}_r , d^{\beta B}_s\}  &=&-\frac{c}{6}\epsilon^{\alpha\beta}\epsilon^{AB}\delta_{r+s,0}
\eea
\bea\label{app com current a d}
[L_m,\alpha_{A\dot{A},n}] &=&-n\alpha_{A\dot{A},m+n} ~~~~~~~[L_m ,d^{\alpha A}_r] =-({m\over2}+r)d^{\alpha A}_{m+r}\cr
\lbrace G^{\alpha}_{\dot{A},r} ,  d^{\beta B}_{s} \rbrace&=&i\epsilon^{\alpha\beta}\epsilon^{AB}\alpha_{A\dot{A},r+s}~~~~~~~
[G^{\alpha}_{\dot{A},r} , \alpha_{B \dot{B},m}]=  -im\epsilon_{AB}\epsilon_{\dot{A}\dot{B}}d^{\alpha A}_{r+m}\cr
[J^a_m,d^{\alpha A}_r] &=&{1\over 2}(\sigma^{Ta})^{\alpha}_{\beta}d^{\beta A}_{m+r}
\eea
\bea\label{app com currents}
[L_m,L_n] &=& {c\over12}m(m^2-1)\delta_{m+n,0}+ (m-n)L_{m+n}\cr
[J^a_{m},J^b_{n}] &=&{c\over12}m\delta^{ab}\delta_{m+n,0} +  i\epsilon^{ab}_{\,\,\,\,c}J^c_{m+n}\cr
\lbrace G^{\alpha}_{\dot{A},r} , G^{\beta}_{\dot{B},s} \rbrace&=&  \epsilon_{\dot{A}\dot{B}}\bigg[\epsilon^{\alpha\beta}{c\over6}(r^2-{1\over4})\delta_{r+s,0}  + (\sigma^{aT})^{\alpha}_{\gamma}\epsilon^{\gamma\beta}(r-s)J^a_{r+s}  + \epsilon^{\alpha\beta}L_{r+s}  \bigg]\cr
[J^a_{m},G^{\alpha}_{\dot{A},r}] &=&{1\over2}(\sigma^{aT})^{\alpha}_{\beta} G^{\beta}_{\dot{A},m+r}\cr
[L_{m},J^a_n]&=& -nJ^a_{m+n}\cr
[L_{m},G^{\alpha}_{\dot{A},r}] &=& ({m\over2}  -r)G^{\alpha}_{\dot{A},m+r}
\eea
We define $J^{+}_n, J^-_n$ as
\bea
J^+_n &=& J^1_n + i J^2_n\cr
J^-_n&=& J^1_n - i J^2_n
\eea
From (\ref{app com current a d}), one can see that $d^{\alpha A}_{n}$ with $\alpha=+,-$ is a $SU(2)_L$ charge doublet. We have
\bea
[J^{+}_m,d^{+ A}_r] &=& 0,\qquad~~~~~ [J^{-}_m,d^{+ A}_r] ~=~ d^{-A}_{m+r}\cr
[J^{-}_m,d^{+ A}_r] &=& d^{-A}_{m+r},\qquad [J^{+}_m,d^{+ A}_r] ~=~ 0
\eea
From (\ref{app com currents}), one can see that $G^{\alpha}_{\dot{A},r}$  with $\alpha=+,-$ is also a $SU(2)_L$ charge doublet. We have
\bea
[J^{+}_{m},G^{+}_{\dot{A},r}]  &=& 0 ,\qquad\qquad ~~~[J^{-}_{m},G^{+}_{\dot{A},r}]  ~=~ G^{-}_{\dot{A},m+r}\cr
[J^{+}_{m},G^{-}_{\dot{A},r}]  &=&G^{+}_{\dot{A},m+r},\qquad ~[J^{-}_{m},G^{-}_{\dot{A},r}]  ~=~ 0 
\eea
It is believed that the contracted large $\mathcal N=4$ superconformal algebra is an exact symmetry at any point of the moduli space.

Now let's consider the orbifold point.
Look at the winding sector $N=(n_1,n_2,...,n_i,...)$
with the total winding $n=\sum_{i} n_i$. For the $i$th twisted set of copies  with winding number $n_{i}$,
we have following mode expansions on the cylinder.
\bea\label{i string modes}
\alpha^{(i)}_{A \dot A,n}=\frac{1}{2\pi}\int_{\sigma=0}^{2\pi n_{i}}\p_{w}X^{(i)}_{A \dot A}(w)e^{nw}dw\nn
d^{\alpha A (i)}_{n}=\frac{1}{2\pi i}\int_{\sigma=0}^{2\pi n_{i}}\psi^{\alpha A(i)}(w)e^{nw}dw
\eea
In terms of $\alpha$ and $d$ modes, the $J$, $G$ and $L$ modes in (\ref{i string modes}) can be written as
\bea\label{i string modes J G L}
J^{a(i)}_m &=& {1\over 4 n_{i}}\sum_{r}\epsilon_{AB}d^ {\gamma B(i)}_r\epsilon_{\alpha\gamma}(\sigma^{aT})^{\alpha}_{\beta}d^ {\beta A(i)}_{m-r},\qquad a=1,2,3\cr
J^{3(i)}_m &=&  - {1\over 2 n_{i}}\sum_{r} d^ {+ +(i)}_{r}d^ {- -(i)}_{m-r} - {1\over 2 n_{i}}\sum_{r}d^ {- +(i)}_r d^ {+ -(i)}_{m-r}\cr
J^{+(i)}_m&=&\frac{1}{n_{i}}\sum_{r}d^ {+ +(i)}_rd^ {+ -(i)}_{m-r} ,\qquad J^{-(i)}_m=\frac{1}{n_{i}}\sum_{r}d^ {--(i)}_rd^ {- +(i)}_{m-r}\cr
G^{\alpha(i)}_{\dot{A},r} &=& -\frac{i}{n_{i}}\sum_{n}d^ {\alpha A(i)}_{r-n} \alpha^{(i)}_{A\dot{A},n}\cr
L^{(i)}_m&=& -{1\over 2 n_i}\sum_{n} \epsilon^{AB}\epsilon^{\dot A \dot B}\alpha^{(i)}_{A\dot{A},n}\alpha^{(i)}_{B\dot{B},m-n}- {1\over 2 n_i}\sum_{r}(m-r+{1\over2})\epsilon_{\alpha\beta}\epsilon_{AB}d^ {\alpha A(i)}_r d^ {\beta B(i)}_{m-r}
\eea
Let $k$ be an integer. 
The mode numbers for $\alpha,L,J$ are $n=k/n_{i}$.
In the R sector, the mode numbers for $d$ and $G$ are $n=k/n_{i}$.
In the NS sector, the mode numbers for $d$ and $G$ are $n=(k+\frac{1}{2})/n_{i}$.
The modes (\ref{i string modes})  and (\ref{i string modes J G L}) satisfy the contracted large $\mathcal N=4$ superconformal algebra (\ref{app com a d})(\ref{app com current a d})(\ref{app com currents}) with $c=6n_{i}$. 

We define the global modes $O^{g}_n$ by summing the terms from each copy
\be\label{global modes app}
O^{g}_n=\sum_{i}O^{(i)}_n
\ee
where the modes $O$ can be modes of $\alpha~d~L~J~G$.
The global modes satisfy the contracted large $\mathcal N=4$ superconformal algebra (\ref{app com a d})(\ref{app com current a d})(\ref{app com currents}) with $c=6n$. It is believed that global modes satisfy the algebra at any point in the moduli space.

\section{Hermitian conjugation}\label{app conjugate}
We use the following Hermitian conjugation rules
\bea
( G^{+}_{+}(\tau,\sigma))^{\dagger}=-G^{-}_{-}(-\tau,\sigma)
~~~~~~~
( G^{+}_{-}(\tau,\sigma))^{\dagger}=G^{-}_{+}(-\tau,\sigma)\nn
( \bar G^{+}_{+}(\tau,\sigma))^{\dagger}=-\bar G^{-}_{-}(-\tau,\sigma)
~~~~~~~
( \bar G^{+}_{-}(\tau,\sigma))^{\dagger}=\bar G^{-}_{+}(-\tau,\sigma)
\eea
and 
\be
(\sigma^{--}(\tau,\sigma))^{\dagger}=-\sigma^{++}(-\tau,\sigma)
~~~~~~~
(\sigma^{-+}(\tau,\sigma))^{\dagger}=\sigma^{+-}(-\tau,\sigma)
\ee
In this convention, the deformation operator is a Hermitian operator
\bea
(D(\tau,\sigma))^{\dagger}=D(-\tau,\sigma)
\eea
From the definition of $\bar G^{(P)}$ (\ref{GN p}), we find
\be
\bar G^{+(P)\dagger}_{+,0}=-\bar G^{-(P)}_{-,0}~~~~~~
\bar G^{+(P)\dagger}_{-,0}=\bar G^{-(P)}_{+,0}
\ee
We also have the useful relations 
\bea\label{A3}
 G_{\dot{A},-\frac{1}{2}}^{+(0)}\sigma^{ + \bar\alpha}=0~~~~~~~~ G_{\dot{A},-\frac{1}{2}}^{-(0)}\sigma^{ -\bar \alpha}=0\nn
\bar G_{\dot{A},-\frac{1}{2}}^{+(0)}\sigma^{ \alpha +}=0~~~~~~~~\bar G_{\dot{A},-\frac{1}{2}}^{-(0)}\sigma^{ \alpha -}=0
\eea
and
\be\label{A4}
G_{\dot{A},-\frac{1}{2}}^{-(0)}\sigma^{ + \bar\alpha}=-G_{\dot{A},-\frac{1}{2}}^{+(0)}\sigma^{ - \bar\alpha}
~~~~~~
\bar G_{\dot{A},-\frac{1}{2}}^{-(0)}\sigma^{ \alpha +}=-\bar G_{\dot{A},-\frac{1}{2}}^{+(0)}\sigma^{ \alpha -}
\ee

\section{Global zero mode characters}\label{app chara}

In this appendix, we will find the characters for the left moving subalgebra formed by $d_{0}^{\alpha A}$, $J^{a}_{0}$ and $G^{\alpha}_{\dot A}$.
We will also find the characters for right moving subalgebra formed by $\bar d^{\bar \alpha A}_0$.

\subsection{Left sector}

Consider the subalgebra formed by the $d_{0}^{\alpha A}$, $J^{a}_{0}$ and $G^{\alpha}_{\dot A}$. The lowest weight states $\phi$ of the irreducible representations are defined by
\be\label{A lowest states}
d^{-A}_{0}|\phi\rangle=G^{-}_{\dot A,0}|\phi\rangle=J^{-}_{0}|\phi\rangle=0
\ee
$\phi$ carries  the  charge
\be\label{A lowest s charge}
J^{3}_{0}|\phi\rangle=-j|\phi\rangle
\ee

To find the irreducible representations of the subalgebra, we use the Racah-Speiser algorithm \cite{Cordova:2016emh,Fuchs:1997jv}. Start with a lowest weight state defined by (\ref{A lowest states}), with charge $-j$. The modes $d_{0}^{+ A}$ and $G^{+}_{\dot A,0}$ are four fermionic raising operators.  Applying these raising operators generates a set of states with charges  $j^{3}=-j,j+1/2,\ldots, -j+2$. One may now try to apply  $J^{+}_{0}$ to each of these states, filling out the $J^a_0$ representations. Such  a $J^a_0$ multiplet with highest weight $j$ will be denoted by $[j]$. Thus formally, we get multiplets of the type  $[j],[j-1/2],\ldots,[j-2]$.

The Racah-Speiser algorithm says the following. For  $j\geq 0$ we can indeed have such a multiplet $[j]$. But there cannot be multiplets $[j]$ with $j<0$. For the $SU(2)$ algebra, we have the following rules. The multiplet  $[-1/2]$ does not correspond to an actual multiplet.   A multiplet $[j]$ with $j\leq -1$ cancels a multiplet $[-j-1]$. For example, a $[-1]$ cancels a $[0]$ and a $[-3/2]$ cancels a $[1/2]$.

The character is defined by a trace over the irreducible representation
\be
\chi^{L}_{j}(y)={\rm Tr} (-1)^{2J^{3}_{0}} y^{2J^{3}_{0}}
\ee
Let's find the characters for the irreducible representations with $j\ge 2$. There are four pairs of fermionic raising and lowering operators, thus an irreducible representation contains the following $SU(2)$ multiplets 
\be
[j] \rightarrow 4[j-1/2] \rightarrow 6[j-1] \rightarrow 4[j-3/2] \rightarrow [j-2]
\ee
where $\rightarrow$ corresponds to  one of the fermionic raising operators and $n[j]$ means there are $n$ of $[j]$ multiplets. Thus the character of such a left sector multiplet is 
\bea\label{global character left}
\chi^{L}_{j}(y)&=&\chi_{j}(-y)+4\chi_{j-1/2}(-y)+6\chi_{j-1}(-y)+4\chi_{j-3/2}(-y)+\chi_{j-2}(-y)\nn
&=&\chi_{j-1}(-y)(y^{1/2}-y^{-1/2})^4
\eea
where $\chi_{j}(-y)$ is the character for a SU(2) multiplet $[j]$
\be
\chi_{j}(-y)=(-1)^{2j}(y^{2j}+y^{2j-2}+\ldots+ y^{-2j})=(-1)^{2j}\frac{y^{2j+1}-y^{-2j-1}}{y-y^{-1}}
\ee

We now  check the special cases  $j=3/2$ and $j=1$. We will see that they also yield the expression  (\ref{global character left}). For $j=1$, by applying the four fermion operators, we get
\be
[1] \rightarrow 4[1/2] \rightarrow 6[0] \rightarrow 4[-1/2] \rightarrow [-1]
\ee
Follow the Racah-Speiser algorithm, we exclude the $4[-1/2]$ and the $[-1]$ cancels one of the $[0]$ multiplets. Thus we left with
\be
[1],~~ 4[1/2],~~ 5[0]
\ee
The character is
\be
\chi^{L}_{1}(y)=\chi_{1}(-y)+4\chi_{1/2}(-y)+5\chi_{0}(-y)=(y^{1/2}-y^{-1/2})^4
\ee
which agrees with the expression (\ref{global character left}). For $j=3/2$,  applying the four fermion raising operators gives
\be
[3/2] \rightarrow 4[1] \rightarrow 6[1/2] \rightarrow 4[0] \rightarrow [-1/2]
\ee
Follow the Racah-Speiser algorithm, we exclude the $[-1/2]$. Thus we are left with
\be
[3/2],~~ 4[1],~~ 6[1/2],~~4[0]
\ee
The character is
\be
\chi^{L}_{3/2}(y)=\chi_{3/2}(-y)+4\chi_{1}(-y)+6\chi_{1/2}(-y)+4\chi_{0}(-y)=-(y+y^{-1})(y^{1/2}-y^{-1/2})^4
\ee
This again satisfies (\ref{global character left}). Thus (\ref{global character left}) is correct for all $j\ge 1$ which are the cases that we will need.

\subsection{Right sector}

For the right movers, we consider the sub-algebra formed by $\bar d^{\bar\alpha A}_0$.
The lowest weight states 
are defined by
\be
\bar d^{- A}_{0}| \phi\rangle=0
\ee
with  charge 
\be
\bar J^{3}_{0}| \phi\rangle=\bar j_{3} | \phi\rangle
\ee
There are two fermionic raising operators. Application of these to a lowest weight state with charge $\bar j_3$ gives  states with charges
\be
\bar j_{3} \rightarrow 2(\bar j_{3}+1/2) \rightarrow \bar j_{3}+1 
\ee
The character is defined by the trace over the irreducible representation. We find
\be
\chi^{R}_{\bar j_{3}}(\bar y)={\rm Tr} (-1)^{-2\bar J^{3}_{0}} \bar y^{2\bar J^{3}_{0}}
=-(-\bar y)^{2\bar j_{3}+1}(\bar y^{1/2}-\bar y^{-1/2})^2
\ee

\section{Global states are unlifted}\label{app global unlifted}

In this appendix, we will show that global states from the contracted large $\mathcal{N}=4$ superconformal algebra are unlifted. 
Global states are states with global modes acting on the Ramond ground states.

The deformation operator (\ref{D 1/4}) preserves the $\mathcal{N}=4$ superconformal symmetry. The generators of the $\mathcal{N}=4$ superconformal symmetry are global modes $J^{a}_{n}$, $G^{\alpha}_{\dot A,r}$ and $L_n$. Thus these global modes do not change the lift. Since the Ramond ground states are unlifted, global states obtained by  acting with global modes $J^{a}_{n}$, $G^{\alpha}_{\dot A,r}$ and $L_n$ to the Ramond ground states are unlifted.
  
The global modes $\alpha_{A\dot A,n}$ and $d^{\alpha A}_{r}$ are extra generators in the contracted large $\mathcal{N}=4$ superconformal algebra.
It is not obvious that the global modes $\alpha_{A\dot A,n}$ and $d^{\alpha A}_{r}$ do not change the lift. In the following we will show that global modes $\alpha_{A\dot A,n}$ and $d^{\alpha A}_{r}$ commute with the deformation operator (\ref{D 1/4}). Thus they also do not contribute to the lift. 
Hence all the global states from the contracted large $\mathcal{N}=4$ algebra are unlifted.

The deformation operator (\ref{D 1/4}) has the left moving part $G^{+}_{\dot B, -\frac{1}{2}}\sigma^{-}$.
Let us consider

\bea\label{d D}
\{d^{\alpha A}_{n}, G^{+}_{\dot B, -\frac{1}{2}}\sigma^{-}(0)\}
&=& \oint_{C_{0}}\frac{dw}{2\pi i} \psi^{\alpha A}(w)e^{nw}G^{+}_{\dot B, -\frac{1}{2}}\sigma^{-}(0)\nn
&=&\oint_{C_{0}}\frac{dw}{2\pi i} \psi^{\alpha A}(w)(1+nw+\dots)G^{+}_{\dot B, -\frac{1}{2}}\sigma^{-}(0)\nn
&=& d^{\alpha A}_{\frac{1}{2}}G^{+}_{\dot B, -\frac{1}{2}}\sigma^{-}(0)
\eea
where higher terms in the expansion vanish. From the commutator in appendix \ref{commutators},
we have
\bea
\{ d^{\alpha A}_{\frac{1}{2}}, G^{+}_{\dot{B},-\frac{1}{2}} \}&=&i\epsilon^{+\alpha}\epsilon^{BA}\alpha_{B\dot{B},0}
\eea
Thus  the last expression in (\ref{d D}) becomes
\be \label{first d D}
d^{\alpha A}_{\frac{1}{2}}G^{+}_{\dot B, -\frac{1}{2}}\sigma^{-}(0)
=i\epsilon^{+\alpha}\epsilon^{BA}\alpha_{B\dot{B},0}\sigma^{-}(0)-G^{+}_{\dot B, -\frac{1}{2}}d^{\alpha A}_{\frac{1}{2}}\sigma^{-}(0)
\ee
Note that  $\alpha_{B\dot{B},0}\sigma^{-}(0)=0$ since $\sigma^{-}(0)$ has no momentum along $T^4$. Also, the
operator $d^{\alpha A}_{\frac{1}{2}}\sigma^{-}(0)$ must vanish as it has dimension $0$ but carries a nontrivial charge.  Thus we find
\bea\label{d D zero}
\{d^{\alpha A}_{n}, G^{+}_{\dot B, -\frac{1}{2}}\sigma^{-}(0)\}
=0
\eea

In a similar manner, we have
\be
[\alpha_{A\dot A,n}, G^{+}_{\dot B, -\frac{1}{2}}\sigma^{-}(0)]
=\alpha_{A\dot A,0}G^{+}_{\dot B, -\frac{1}{2}}\sigma^{-}(0)+n\alpha_{A\dot A,1}G^{+}_{\dot B, -\frac{1}{2}}\sigma^{-}(0)
\ee
The first term is zero since the operator $G^{+}_{\dot B, -\frac{1}{2}}\sigma^{-}(0)$ has no momentum along $T^4$. The second term is zero since it has dimension $0$ but carries a nonzero  charge $A$.  Thus we find 
\be
[\alpha_{A\dot A,n}, G^{+}_{\dot B, -\frac{1}{2}}\sigma^{-}(0)]
=0
\ee

\section{Some properties of left and right movers}\label{app prop r and l}

In this appendix, we will note some properties which will be of use to constructing the long multiplets in section \ref{section lifting}.

\subsection{Right moving sector}\label{app prop r}

In this appendix, we will find some properties of right moving states discussed in section \ref{s right mover}.

First let us show that having the right moving sector state (\ref{r mover zero lifting}) leads to zero lift.
Applying  $\mathcal P\bar \sigma^{+}$ to (\ref{r mover zero lifting}) and using the relations in section \ref{s twist}, we find
\bea
&&\mathcal P\bar \sigma^{+}(\bar d_0^{+A(1)}-\bar d_0^{+A(2)})|\bar 0^{-}_{R}\rangle|\bar 0^{-}_{R}\rangle\nn
&=&\mathcal P\left(\sum_{p}f^{F+}_{a}[0,p]\bar d^{+A}_{p/2}-\sum_{p}f^{F+}_{-a}[0,p]\bar d^{+A}_{p/2}\right)|\bar\chi\rangle\nn
&=&\mathcal P\left[\left(\frac{1}{2}\bar d^{+A}_{0}+\frac{i}{2}a\bar d^{+A}_{-1/2}+O(a^3)\right)-\left(\frac{1}{2}\bar d^{+A}_{0}-\frac{i}{2}a\bar d^{+A}_{-1/2}+O(a^3)\right)\right]\left(|\bar 0^{2-}_{R}\rangle+O(a)\right)\nn
&=&0
\eea
In the last step, we have used the observation after eq.\,(\ref{f+ 0}) that taking the projection onto the right moving Ramond ground state is equivalent to taking the limit $a\rightarrow 0$.
Applying $\mathcal P\bar \sigma^{-}$ to the same state (\ref{r mover zero lifting}), we find
\bea
&&\mathcal P\bar \sigma^{-}(\bar d_0^{+A(1)}-\bar d_0^{+A(2)})|\bar 0^{-}_{R}\rangle|\bar 0^{-}_{R}\rangle\nn
&=&[\bar J^{-}_{0},\mathcal P\bar \sigma^{+}](\bar d_0^{+A(1)}-\bar d_0^{+A(2)})|\bar 0^{-}_{R}\rangle|\bar 0^{-}_{R}\rangle\nn
&=&\bar J^{-}_{0}\mathcal P\bar \sigma^{+}(\bar d_0^{+A(1)}-\bar d_0^{+A(2)})|\bar 0^{-}_{R}\rangle|\bar 0^{-}_{R}\rangle-\mathcal P\bar \sigma^{+}\bar J^{-}_{0}(\bar d_0^{+A(1)}-\bar d_0^{+A(2)})|\bar 0^{-}_{R}\rangle|\bar 0^{-}_{R}\rangle\nn
&=&0
\eea
Thus any state of the system which has the state (\ref{r mover zero lifting}) as its right moving component will have zero lift. 

Now let us check the long multiplet structure (\ref{right mover}).
First we note that
\be\label{anni bottom}
\mathcal P\bar \sigma^{-} |\bar 0^{-}_{R}\rangle|\bar 0^{-}_{R}\rangle=0
\ee
since there is no right Ramond ground state in doubly wound sector with $\bar j<-1/2$. From the second diagram in \ref{multiplets diagram} we see that $\phi^{R}$ should be the right moving state for the bottom member of the long multiplet. (There were only $4$ states in the right sector and we have already seen that (\ref{r mover zero lifting}) cannot be the right state for a state in the long multiplet. In the following we can see that the other states in (\ref{right mover}) are not annihilated by $\mathcal P\bar \sigma^{-}$, so $\phi^{R}$ is the unique choice for the bottom member of the multiplet.) 

Next, using  (\ref{chi}) for the right sector we find
\be\label{p move bottom}
\mathcal P\bar \sigma^{+} |\bar 0^{-}_{R}\rangle|\bar 0^{-}_{R}\rangle=|\bar 0^{2-}_{R}\rangle
\ee
This shows that  $\phi^{R}_{+}$ and $\phi^{R}_{-}$ in (\ref{right mover}) are the right sector states for the middle members of the long multiplet.

Finally, we will show that 
\be\label{r top}
\mathcal P\bar \sigma^{+} |\bar 0^{2-}_{R}\rangle
=\frac{1}{2}(\bar d_0^{++(1)}-\bar d_0^{++(2)})(\bar d_0^{+-(1)}-\bar d_0^{+-(2)})|\bar 0^{-}_{R}\rangle|\bar 0^{-}_{R}\rangle
\ee
This will confirm that $\phi^{R}_{+-}$ in (\ref{right mover}) is indeed the right moving state for the top member of the long multiplet. 
First we note that the global modes $\bar d^{\bar \alpha A}_0$ commute with the operator $\bar \sigma^{\bar \beta}$
\bea\label{right d zeromode sigma}
[\bar d^{\bar \alpha A}_{0}, \bar \sigma^{\bar \beta}(0)]
= \oint_{C_{0}}\frac{ d\bar w}{2\pi i}\bar \psi^{\bar \alpha A}(\bar w)\bar\sigma^{\bar\beta}(0) 
= \bar d^{\bar \alpha A}_{-1/2}\bar \sigma^{\bar\beta}(0)=0
\eea
The last equality follows  because there is no operator that has dimension $\bar h=0$ and a nonzero charge $A$.
Then we have
\be\label{ddp 1}
\bar d_0^{++}\bar d_0^{+-}\mathcal P\bar \sigma^{+} |\bar 0^{2-}_{R}\rangle=\mathcal P\bar \sigma^{+}\bar d_0^{++}\bar d_0^{+-} |\bar 0^{2-}_{R}\rangle
\ee
where $\bar d^{\bar \alpha A}_0$ is the global mode.
Using
\be
\bar d_0^{++}\bar d_0^{+-} |\bar 0^{2-}_{R}\rangle = 2 |\bar 0^{2+}_{R}\rangle
\ee
and
\be
\mathcal P\bar \sigma^{+}|\bar 0^{2+}_{R}\rangle=|\bar 0^{+}_{R}\rangle|\bar 0^{+}_{R}\rangle
\ee
(\ref{ddp 1}) becomes
\be\label{ddp}
\bar d_0^{++}\bar d_0^{+-}\mathcal P\bar \sigma^{+} |\bar 0^{2-}_{R}\rangle
=2|\bar 0^{+}_{R}\rangle|\bar 0^{+}_{R}\rangle
\ee
Since the global modes $\bar d^{-  A}_0$ commute with the $\mathcal P\bar \sigma^{+}$, we have
\be
\bar d_0^{-  A}\mathcal P\bar \sigma^{+} |\bar 0^{2-}_{R}\rangle
=\mathcal P\bar \sigma^{+}\bar d_0^{-  A}|\bar 0^{2-}_{R}\rangle
=0
\ee
We now apply $\bar d_0^{--}\bar d_0^{-+}$ to both sides of (\ref{ddp}). This gives
\be
|\phi^{R}_{+-}\rangle=\mathcal P\bar \sigma^{+} |\bar 0^{2-}_{R}\rangle
=\frac{1}{2}\bar d_0^{--}\bar d_0^{-+}|\bar 0^{+}_{R}\rangle|\bar 0^{+}_{R}\rangle
\label{zztwo}
\ee
Now we write 
\bea
|\bar 0^{+}_{R}\rangle|\bar 0^{+}_{R}\rangle=\bar d_0^{++(1)} \bar d_0^{+-(1)} \bar d_0^{++(2)} \bar d_0^{+-(2)}|\bar 0^{-}_{R}\rangle|\bar 0^{-}_{R}\rangle
\eea
Using this on the RHS of (\ref{zztwo}) we get (\ref{r top}). 
Since there is no right Ramond ground state in doubly wound sector with $\bar j>1/2$, we have
\be\label{top right anni}
\mathcal P\bar \sigma^{+}|\bar 0^{+}_{R}\rangle|\bar 0^{+}_{R}\rangle=0
\ee
We can see that (\ref{zztwo}) cannot be raised by $\mathcal P\bar \sigma^{+}$
\be\label{anni top}
\mathcal P\bar \sigma^{+}|\phi^{R}_{+-}\rangle
=\frac{1}{2}\mathcal P\bar \sigma^{+} \bar d_0^{--}\bar d_0^{-+}|\bar 0^{+}_{R}\rangle|\bar 0^{+}_{R}\rangle
=\frac{1}{2}\bar d_0^{--}\bar d_0^{-+}\mathcal P\bar \sigma^{+}|\bar 0^{+}_{R}\rangle|\bar 0^{+}_{R}\rangle=0
\ee
Thus $|\phi^{R}_{+-}\rangle$ is indeed the top member of the long multiplet.

 Now let us establish the properties given in the right diagram in (\ref{multiplets diagram right}).
Starting from (\ref{zztwo}), applying $\mathcal P\bar \sigma^{-}$ gives
\be\label{right diag prop 1}
\mathcal P\bar \sigma^{-}|\phi^{R}_{+-}\rangle
=\frac{1}{2}\mathcal P\bar \sigma^{-} \bar d_0^{--}\bar d_0^{-+}|\bar 0^{+}_{R}\rangle|\bar 0^{+}_{R}\rangle
=\frac{1}{2}\bar d_0^{--}\bar d_0^{-+}\mathcal P\bar \sigma^{-}|\bar 0^{+}_{R}\rangle|\bar 0^{+}_{R}\rangle
=\frac{1}{2}\bar d_0^{--}\bar d_0^{-+}|\bar 0^{2+}_{R}\rangle
=|\bar 0^{2-}_{R}\rangle
\ee
Apply $\mathcal P\bar \sigma^{-}$ again we have
\be\label{right diag prop 2}
\mathcal P\bar \sigma^{-} |\bar 0^{2-}_{R}\rangle=|\bar 0^{-}_{R}\rangle|\bar 0^{-}_{R}\rangle
\ee
These relations verify the relations in the right diagram of (\ref{multiplets diagram right}).

\subsection{Left mover}\label{app prop l}

In this appendix, we will verify the relation (\ref{left amplitude}) used  in section \ref{s left mover}.

Using  property (\ref{A4}) in Appendix \ref{app conjugate}, the left part of the operator $\bar G^{\bar \alpha (P)}_{\dot A,0}$ (\ref{GN p s}) can be written as
\be
\pi \mathcal P G^{+}_{\dot A,-\frac{1}{2}}\sigma^{-}=-\pi \mathcal P G^{-}_{\dot A,-\frac{1}{2}}\sigma^{+}
\ee
Thus (\ref{left amplitude}) is equivalent to 
\bea\label{left amplitude app}
\mathcal P G^{-}_{\dot A,-\frac{1}{2}}\sigma^{+}(d_{-1}^{--(1)}-d_{-1}^{--(2)})(d_0^{+-(1)}-d_0^{+-(2)})|0^{-}_{R}\rangle|0^{-}_{R}\rangle=i d^{--}_{-1/2}\alpha_{+\dot A,-1/2}|0^{2-}_{R}\rangle
\eea
This relation  (\ref{left amplitude app}) is obtained as follows. Using the relations in section \ref{s twist}, we have
\bea
&&\sigma^{+}d^{--(1)}_{-1}d_0^{+-(1)}|0^{-}_{R}\rangle|0^{-}_{R}\rangle\nn
&=&(\sum_{p}f^{F-}_{a}[-1,p]d^{--}_{p/2})(\sum_{q}f^{F+}_{a}[0,q]d^{+-}_{q/2})|\chi\rangle\nn
&=&\left(-\frac{i}{2}a^{-1}d^{--}_{-1/2}+\frac{1}{2}d^{--}_{-1}+O(a)\right)\left(\frac{1}{2}d^{+-}_{0}+\frac{i}{2}a d^{+-}_{-1/2}+O(a^3)\right)|\chi\rangle
\eea
and 
\bea
&&\sigma^{+}d^{--(1)}_{-1}d_0^{+-(2)}|0^{-}_{R}\rangle|0^{-}_{R}\rangle\nn
&=&(\sum_{p}f^{F-}_{a}[-1,p]d^{--}_{p/2})(\sum_{q}f^{F+}_{-a}[0,q]d^{+-}_{q/2})|\chi\rangle\nn
&=&\left(-\frac{i}{2}a^{-1}d^{--}_{-1/2}+\frac{1}{2}d^{--}_{-1}+O(a)\right)\left(\frac{1}{2}d^{+-}_{0}-\frac{i}{2}a d^{+-}_{-1/2}+O(a^3)\right)|\chi\rangle
\eea
Then we get
\bea
&&\mathcal P \sigma^{+}d_{-1/2}^{--(1)}(d_0^{+-(1)}-d_0^{+-(2)})|0^{-}_{R}\rangle|0^{-}_{R}\rangle\nn
&=&\mathcal P \left(-\frac{i}{2}a^{-1}d^{--}_{-1/2}+\frac{1}{2}d^{--}_{-1}+O(a)\right)\left(i a d^{+-}_{-1/2}+O(a^3)\right)|\chi\rangle\nn
&=&\frac{1}{2}d^{--}_{-1/2}d^{+-}_{-1/2}|0^{2-}_{R}\rangle
\eea
By using the commutators in Appendix \ref{commutators}
\bea\label{com for lift l}
[G^{-}_{\dot A,0},d^{+-}_{n}]=-i\alpha_{+\dot A,n}~~~~[G^{-}_{\dot A,0},d^{--}_{n}]=0
\eea
we have
\bea\label{lift left rela 1}
&&\mathcal P G^{-}_{\dot A,0}\sigma^{+}d_{-1}^{--(1)}(d_0^{+-(1)}-d_0^{+-(2)})|0^{-}_{R}\rangle|0^{-}_{R}\rangle\nn
&=&G^{-}_{\dot A,0}\frac{1}{2}d^{--}_{-1/2}d^{+-}_{-1/2}|0^{2-}_{R}\rangle
=\frac{i}{2}d^{--}_{-1/2}\alpha_{+\dot A,-1/2}|0^{2-}_{R}\rangle
\eea
From (\ref{com for lift l}) and that $G^{-}_{\dot A,0}$ annihilates all the Ramond ground states including $(d_0^{+-(1)}-d_0^{+-(2)})|0^{-}_{R}\rangle|0^{-}_{R}\rangle$, we have
\be\label{lift left rela 2}
G^{-}_{\dot A,0}d_{-1}^{--(1)}(d_0^{+-(1)}-d_0^{+-(2)})|0^{-}_{R}\rangle|0^{-}_{R}\rangle=0
\ee

To get (\ref{left amplitude app}), writing  $G^{-}_{\dot A, -\h}$ as a contour operator surrounding  $\sigma^{+}$ and breaking the contour into two zero modes wrapping the cylinder, one above and one below the point $w$:
\bea
G^{-}_{\dot A,-\frac{1}{2}}\sigma^{+}(w)&=&\oint_{c_w} \frac{dw'}{2\pi i} G^{-}_{\dot A}(w')\sigma^{+}(w)\nn
&=&\int_{0}^{2\pi} \frac{d\sigma}{2\pi} G^{-}_{\dot A}(\tau>\tau_{w},\sigma)\sigma^{+}(w)
-\sigma^{+}(w)\int_{0}^{2\pi} \frac{d\sigma}{2\pi} G^{-}_{\dot A}(\tau<\tau_w,\sigma)\nn
&=& G^{-}_{\dot A,0}(\tau>\tau_w)\sigma^{+}(w)-\sigma^{+}(w)G^{-}_{\dot A,0}(\tau<\tau_w)
\eea
Then we get
\bea\label{left cal copy 1}
&&\mathcal P G^{-}_{\dot A,-\frac{1}{2}}\sigma^{+}d_{-1}^{--(1)}(d_0^{+-(1)}-d_0^{+-(2)})|0^{-}_{R}\rangle|0^{-}_{R}\rangle\nn
&=&\mathcal P \left[G^{-}_{\dot A,0}\sigma^{+}d_{-1}^{--(1)}(d_0^{+-(1)}-d_0^{+-(2)})
-\sigma^{+}G^{-}_{\dot A,0}d_{-1}^{--(1)}(d_0^{+-(1)}-d_0^{+-(2)})\right]|0^{-}_{R}\rangle|0^{-}_{R}\rangle\nn
&=&\frac{i}{2}d^{--}_{-1/2}\alpha_{+\dot A,-1/2}|0^{2-}_{R}\rangle
\eea
where in the last step we use (\ref{lift left rela 1}) and (\ref{lift left rela 2}).
By replacing $a\rightarrow -a$, we can exchange the excitations on copy 1 and copy 2. Since (\ref{left cal copy 1}) doesn't depend on $a$, we have
\be\label{left cal copy 2}
\mathcal P G^{-}_{\dot A,-\frac{1}{2}}\sigma^{+}d_{-1}^{--(2)}(d_0^{+-(2)}-d_0^{+-(1)})|0^{-}_{R}\rangle|0^{-}_{R}\rangle
=\frac{i}{2}d^{--}_{-1/2}\alpha_{+\dot A,-1/2}|0^{2-}_{R}\rangle
\ee
Adding (\ref{left cal copy 1}) and (\ref{left cal copy 2}) gives (\ref{left amplitude app}).

\end{document}